\documentclass[nohyper,12pt,letterpaper]{JHEP}  
\newfont{\frak}{eufm10 scaled 1200}
\newcommand{\mfrak}[1]{\mbox{\frak #1}}
\newfont{\Bbb}{msbm10 scaled 1200}     
\newcommand{\mathbb}[1]{\mbox{\Bbb #1}}
\DeclareSymbolFont{AMSa}{U}{msa}{m}{n}
\DeclareSymbolFont{AMSb}{U}{msb}{m}{n}
\let\Box\relax
\DeclareMathSymbol{\Box}{\mathord}{AMSa}{"03} 

\def\IZ{{\mathbb Z}}
\def\IR{{\mathbb R}}

\def\ignorethis#1{}
\def\be{\begin{equation}}
\def\ee{\end{equation}}
\def\bear{\begin{eqnarray}}
\def\eear{\end{eqnarray}}
\def\nn{\nonumber}

\def \tb#1{\left(\begin{array}#1\end{array}\right)}
\def \abs#1{\left|#1\right|}

\def \ket#1{\left\vert #1\right\rangle}
\def \dslash{{\not\!\partial}}

\def \ignoruj#1{}
\def \eqn#1#2{\begin{equation}#2\label{#1}\end{equation}}

\def \asd{anti-self-dual }
\def \asdW{anti-self-dual}

\def \ham{\mathcal H}
\def \Lag{\mathcal L}
\def \action{\mathcal A}

\newcommand\px[1]{{\partial_{#1}}}
\newcommand\qx[1]{{\partial^{#1}}}

\newcommand\rep[1]{{\underline{\bf {#1}}}}      
\newcommand\tr[1]{{\mbox{tr}\{{#1}\}}}          
\newcommand\com[2]{{\lbrack {#1},{#2}\rbrack}}  
\newcommand\acom[2]{{\{ {#1},{#2} \}}}          

\def\lbr{{\lbrack}}
\def\rbr{{\rbrack}}

\def\wdg{{\wedge}}                              

\def\Id{{\bf I}}                                

\def\Imx{{\mbox{Im}}}                           

\def\hacek{\accent20}                           
\def\a{{\alpha}}
\def\b{{\beta}}
\def\g{{\gamma}}
\def\d{{\delta}}
\def\del{{\delta}}
\def\e{{\epsilon}}

\def\u{{\mu}}
\def\v{{\nu}}
\def\r{{\rho}}
\def\s{{\sigma}}
\def\t{{\tau}}

\def\TH{\Theta}
\def\PH{\Phi}
\def\Ga{{\Gamma}}
\def\GPR{{\widetilde{\Gamma}}}

\def\eb{{\bar{\epsilon}}}  
  
\def\bTH{{\bar{\Theta}}}

\newcommand\epd[1]{{\epsilon_{#1}}}              
\newcommand\epu[1]{{\epsilon^{#1}}}              
\newcommand\ept[2]{{{\epsilon_{#1}}^{#2}}}       

\def\bD{{\bar{D}}}                   

\def\sc{{\widetilde{\varphi}}}                  
\def\bsc{{\overline{\widetilde{\varphi}}}}      

\def\CHP{{\Phi}}                     
\def\bCHP{{\overline{\Phi}}}         

\def\vph{{\varphi}}                  

\def\VEV{{\Phi_0}}                   
\def\scVEV{{\varphi_0}}              
\def\Ar{{A}}                         

\def\sig{{{\sigma}}}                 

\def\MR#1{{{\IR}^{#1}}}               
\def\MS#1{{{\bf S}^{#1}}}             
\def\MT#1{{{\bf T}^{#1}}}             
\def\MHT#1{{\hat{{\bf T}}^{#1}}}      

\def\Imx{{\mbox{Im}}}                           

\def\Modsp{{{\cal M}}}                

\newcommand\Ox[1]{{ O({#1}) }}        

\def\da{{\dot{\a}}}

\def\pslash{{\not\! p}}                   
\def\kslash{{\not\! k}}                   

\title{Equations of the (2,0) Theory\\
and Knitted Fivebranes}

\author{Ori Ganor\\
  Department of Physics\\
  Princeton University\\   
  Princeton, NJ 08544\\
E-mail: \email{origa@puhep1.princeton.edu}}

\author{Lubo\hacek s Motl\\
  Department of Physics and Astronomy\\
  Rutgers University\\
  Piscataway, NJ 08855-0849\\
E-mail: \email{motl@physics.rutgers.edu}}

\abstract{
We study non-linear corrections to the low-energy description of the
(2,0) theory. We argue for the existence of a
topological correction term similar to the $C_3\wedge X_8(R)$ in M-theory.
This term can be traced to a classical effect in supergravity
and to a one-loop diagram of the effective 4+1D Super Yang-Mills.
We study other terms which are related to it by supersymmetry and discuss
the requirements on the subleading correction terms from M(atrix)-theory. 
We also speculate on a possible fundamental formulation of the theory.  
}

\keywords{M(atrix) Theories, %
String Duality, Superstring Vacua}

\received{???????? ?st, 1998}
\accepted{???????? ?th, 1998}

\preprint{\hepth{9803108}\\PUPT-1777\\RU-98-07\\HEP-UK-0006}
\begin{document}

%
\noindent\rule\textwidth{.1pt}\vskip 0.5em
\tableofcontents\vskip 0.5em\noindent\rule\textwidth{.1pt}\vskip 0.5em

\section{Introduction}
During the last two years a lot of attention has been devoted to the
newly discovered 5+1D theories \cite{somecom}.
The version of these theories with (2,0) supersymmetry
arises as a low-energy description of type-IIB on an $A_{N-1}$
singularity \cite{somecom} or as the dual low-energy
description of $N$ coincident 5-branes in M-theory \cite{openp}.
Part of the attention
\cite{rozali}-\cite{berozali}
is due to the r\^ole they play in 
compactified M(atrix)-theory \cite{bfss}, part is because they provide
testing grounds to M(atrix)-theory ideas
\cite{prem}-\cite{ABS},
and another part is because they
shed light on non-perturbative phenomena in 3+1D gauge theories
\cite{somecom}.
These theories are also very exciting on their own right. They
lack any parameter which will allow a classical perturbative expansion
(like the coupling constant of SYM).
Thus, these theories have no classical limit (for finite $N$).
the only possible classical expansion is a derivative expansion
where the energy is the small parameter.

One of our goals will be to explore the low-energy description
of the (2,0) theory.
At low energies, and a generic point in moduli space the zeroth order
approximation is $N$ free tensor multiplets which contain the 
chiral \asd 2-forms.
Since the theory contains chiral 2-forms it is more convenient
to write down the low-energy equations of motion rather
than the non-manifestly covariant Lagrangian (there is the
other option of using the manifestly covariant formulation
of \cite{APS,APPS}, but using the equations of motion
will be sufficient for our purposes). These equations 
are to be interpreted \`a la Wilson, i.e. as quantum equations
for operators but with a certain unspecified UV cutoff.
The leading terms in the Wilsonian low-energy description   
are the linear equations of motion for the $N$ free tensor
multiplets. We will be looking for the first sub-leading corrections.
Those corrections will be non-linear and are a consequence of the
interacting nature of the full (2,0) theory.
In general at high enough order in the derivative expansion
the terms in the Wilsonian action are cutoff dependent.
However, we will see that the first order corrections
are independent of the cutoff.   
We will argue that the low-energy equations contain a topological
term somewhat analogous to the subleading $C_3\wdg X_8(R)$ term of
M-theory \cite{VWSCT,DLM}
and which describes a topological correction term
to the \asd string current.
We will then study the implications of supersymmetry.

The paper is organized as follows.
Section (2) is a review of the (2,0) theory.
In section (3) we derive the topological term from the supergravity
limit of $N$ 5-branes of M-theory. Our discussion will be an implementation
of results described in \cite{WTHR}.
In section (4) 
we discuss the implied correction
terms after compactification to 3+1D, and
we find related terms which are implied by supersymmetry.
In section (5) we discuss the currents in 5+1D.
Finally, in section (6-7), we speculate on  a possible ``deeper''
meaning of these correction terms.

After completion of this paper, we received a message about 
related works \cite{Lambert} which studied the single 5-brane solution
in supergravity. We are grateful to N.D. Lambert for the correspondence.


\section{Review of the (2,0) theory}

This section is a short review of some facts we will need
about the (2,0) theory.

\subsection{Realization}

The $(2,0)_N$ theory is realized either as the low-energy decoupled
degrees of freedom from an $A_{N-1}$ singularity (for $N\ge 2$)
of type-IIB \cite{somecom}
or from the low-energy decoupled degrees of freedom of $N$ 5-branes 
of M-theory \cite{openp}.
This is a conformal 5+1D theory which  is interacting for $N>1$.
It has a chiral (2,0) supersymmetry with 16 generators.
One can deform the theory away from the conformal point.
This corresponds to separating the $N$ 5-branes (or blowing
up the $A_{N-1}$ singularity). If the separation scale $x$
is much smaller than the 11D Planck length $M_p^{-1}$ then at energies 
$E\sim M_p^{3/2} x^{1/2}$ one finds a massive decoupled theory
whose low-energy description is given by $N$ free tensor multiplets.

Each free tensor multiplet in 5+1D comprises of 5 scalar fields
$\PH^A$ with $A=1\dots 5$, one tensor field $B_{\u\v}^{(-)}$ where
the $(-)$ indicates that its equations of motion force it to be
\asdW, and 4 multiplets of chiral fermions $\TH$.
The (2,0) supersymmetry in 5+1D has $Sp(2) = Spin(5)_R$ R-symmetry.
The scalars $\PH^A$ are in the $\rep{5}$ whereas the fermions are
in the $(\rep{4},\rep{4})$ of $SO(5,1)\times Sp(2)$ but with
a reality condition. Thus there are 16 real fields in $\TH$.

For the low-energy of the $(2,0)_N$ theory there are $N$ such tensor
multiplets. The moduli space, however, is not just $(\MR{5})^N$
because there are discrete identifications given by the permutation
group. It is in fact $(\MR{5})^N/S_N$. 
Let us discuss what happens for $N=2$. The moduli space can be written
as $\MR{5}\times (\MR{5}/\IZ_2)$. The first $\MR{5}$ is the
sum of the two tensor multiplets. In 5+1D this sum is described
by a free tensor multiplet which decouples from the rest of the theory
(although after compactification, it has some global effects which
do not decouple).
The remaining $\MR{5}/\IZ_2$ is the difference of the two tensor
multiplets. This moduli space has a singularity at the origin
where the low-energy description is no longer two free tensor 
multiplets but is the full conformal theory.

\subsection{Equations of motion for a free tensor multiplet}
To write down the lowest order equations of motion for a free
tensor multiplet we use the field strength
$$
H_{\a\b\g} = 3\px{[\a}B_{\b\g]}^{(-)}.
$$
This equation does not imply that $H$ is \asd but
does imply that $H$ is a closed form. It is possible to modify
this equation such that $H$ will be manifestly \asd.
We will define $H$ to be \asd part of $dB$ according to,
\eqn{hfromb}{H_{\a\b\g}=
\frac 32(\partial_{[\a}B_{\b\g]})
-\frac 14\epsilon_{\a\b\g}{}^{\a'\b'\g'}(\partial_{\a'}B_{\b'\g'}).}
This definition is the same as the previous one for \asd
$dB$, it trivially implies that $H$ is \asd and it
does not lead to the equation $dH=0$ which we will find useful later on.
In any case, we will use the equations of motion for $H$ only
and $B$ will therefore not appear.
For the fermions it is convenient to use 11D Dirac matrices
$$
\Ga^\u,\, \u=0\dots 5,\qquad
\Ga^A,\, A=6\dots 10
$$
with commutation relations
$$
\acom{\Ga^\u}{\Ga^\v} = 2 \eta^{\u\v},\qquad
\acom{\Ga^A}{\Ga^B} = 2\delta^{AB},\qquad
\acom{\Ga^A}{\Ga^\u} = 0.
$$
We define 
$$
\GPR = \Ga^{012345} = \Ga^0\Ga^1\cdots\Ga^5 = \Ga^6\Ga^7\cdots\Ga^{10}
$$
The spinors have positive chirality and satisfy
$$
\TH = \GPR\TH.
$$
The $Spin(5)_R$ acts on $\Ga^A$ while $SO(5,1)$ acts on $\Ga^{\u}$.
The free equations of motion are given by,
\begin{eqnarray}
H^{\u\v\s} &=& {1\over 6}{\epsilon_{\t\r\g}}^{\u\v\s} H^{\t\r\g}
\equiv - \frac16\epsilon^{\u\v\s}{}_{\t\r\g}H^{\t\r\g},\\
\partial_{\lbr\t}{H_{\u\v\s\rbr}} &=& 0,\\
\Box \PH^A &=& 0,\\
\dslash \TH &=& 0.
\end{eqnarray}
The supersymmetry variation is given by,
\begin{eqnarray}
\delta H_{\a\b\g} &=& -\frac{i}{2} \bar\epsilon
                       \Ga_\d\Ga_{\a\b\g}\partial^\d\TH \label{strengvarr} \\
\delta \PH_A      &=& -i \bar\epsilon\Ga_A\TH           \label{phivarr}\\
\delta \TH        &=& (\frac{1}{12} H_{\a\b\g}
                 \Ga^{\a\b\g}+ \Ga^\a\partial_\a
                 \PH_A\Ga^A)\epsilon                    \label{fermivarr}
\end{eqnarray}

The quantization of the theory is slightly tricky.
There is no problem with the  fermions $\TH$ and bosons $\PH^A$,
but the tensor field is self-dual and thus has to be quantized
similarly to a chiral boson in 1+1D.
This means that we second-quantize a free tensor field
without any self-duality constraints and then set to zero
all the oscillators with self-dual polarizations.
The action that we use in 5+1D is:
$$
\action = -{1\over {4\pi}}\int \left\{
 \px{\u}\Phi^A \partial^\u\Phi_A + {3\over 2}\px{[\u}B_{\s\t]}
\partial^{[\u}B^{\s\t]}
 + i 
\bTH
\dslash\TH
 \right\}d^6\s.
$$
Here we have defined $\bTH=\TH^T\Ga^0$.
The normalization is such that integrals of $B_{\s\t}$ over closed
2-cycles 
live on circles of circumference $2\pi$.
In appendix A we list some more useful formulas.

\section{Low-energy correction terms -- derivation from SUGRA}
In this section we will derive a correction term to the zeroth order
low-energy terms.

Let us consider two 5-branes in M-theory.
Let their center of mass be fixed.
The fluctuations of the center
of mass are described by a free tensor multiplet.
Let us assume that the distance between the 5-branes at infinity
$|M_p^{-2}\PH_0|$ is much larger than the 10+1D
Planck length $M_p^{-1}$ and let us consider
the low-energy description of the system for energies
$E\ll |\PH_0|$. The description at lowest order is given by
supergravity in the 10+1D bulk and by a 5+1D tensor multiplet
with moduli space $\MR{5}/\IZ_2$ (we neglect the free tensor
multiplet coming from the overall center of mass).
The lowest order equations of motion for the tensor multiplet are
the same linear equations as described in the previous section.
We would like to ask what are the 
leading nonlinear corrections to the linear equations.

We will now argue that according to the arguments given in \cite{WTHR}
there is a topological contribution to the $dH$ equation of 
motion (here $\PH^{(ij)} \equiv \PH^{(i)} - \PH^{(j)}$)
\begin{equation}\label{eqh}
\px{[\a}H_{\b\g\d]}^{(i)} = 
\sum_{j=1\dots N}^{(j\ne i)}
{3\epu{ABCDE}\over {16\pi |\PH^{(ij)}|^5}}
\PH^{E,(ij)}
\px{[\a}\PH^{A,(ij)}\px{\b}\PH^{B,(ij)}\px{\g}\PH^{C,(ij)}\px{\d]}
\PH^{D,(ij)}.
\end{equation}
Here $A\dots E = 1\dots 5$. $\PH^A$ are the scalars of
the tensor multiplet and $H_{\a\b\g}$ is the \asd field
strength.
Note that the RHS can be written as a pullback $\pi^{*}\omega_4$
of a closed form on the
moduli space which is 
$$
\Modsp \equiv \MR{5}/\IZ_2 - \{0\}.
$$
Here
$$
\pi: \IR^{5,1}\longrightarrow \Modsp = \MR{5}/\IZ_2 - \{0\}
$$
is the map $\PH^A$ from space-time to the moduli space
and,
$$
\omega_4 = {3\over {8\pi^2 |\PH|^5}}
\epu{ABCDE}\PH^E  d\PH^A\wdg d\PH^B\wdg d\PH^C\wdg d\PH^D,
$$
is half an integral form in $H_4(\frac 12\IZ)$, i.e.
$$
\int_{S^4/Z_2} \omega_4 = {1\over 2}.
$$

Let us explain how (\ref{eqh}) arises.
When $\PH^A$ changes smoothly and slowly, the supergravity picture
is that each 5-brane ``wraps'' the other one.
Each 5-brane is a source for the (dual of the) $F_4 = dC_3$ 4-form 
field-strength of 10+1D supergravity. When integrated on a sphere $S^4$
surrounding the 5-brane we get $\int_{S^4} F_4 = 2\pi$.
The other 5-brane now feels an effective $C_3$ flux on its world-volume.
This, in turn, is a source for the 3-form \asd low-energy
field-strength $dH = dC_3$. It follows that the total string charge measured
at infinity of the $\IR^{5,1}$ world-volume of one 5-brane is,
$$
\int dH = \int dC_3 = \int F_4.
$$
The integrals here are on $\IR^4$ which is a subspace of $\IR^{5,1}$
and they measure how much effective string charge passes through
that $\IR^4$. The integral on the RHS can now be calculated.
It is the 4D-angle subtended by the $\IR^4$ relative to the second 
5-brane which was the source of the $F_4$. But this angle can be expressed
solely in terms of $\PH^A$ and the result is the integral over $\omega_4$.


These equations can easily be generalized to $N$ 5-branes.
We have to supplement each field with an index $i=1\dots N$.
We can also argue that there is a correction
\begin{equation}\label{eqphi}
\Box\PH^{D,(i)} = 
-\sum_{j=1\dots N}^{(j\ne i)}
{{\epu{ABCDE}}\over {32\pi |\PH^{(ij)}|^5}}
\PH^{E,(ij)}\px{\a}\PH^{A,(ij)}\px{\b}\PH^{B,(ij)}\px{\g}\PH^{C,(ij)}
H^{\a\b\g,(ij)} + \cdots
\end{equation}
Here $\PH^{(ij)} \equiv \PH^{(i)} - \PH^{(j)}$ and similarly
$H^{(ij)} = H^{(i)} - H^{(j)}$.
The term $(\cdots)$ contains fermions and other contributions.

The equation
(\ref{eqphi}) for $\Box \PH$ can be understood as the equation for force
between a tilted fivebrane and another fivebrane which carries
an $H_{\alpha\beta\gamma}$ flux.
As far as BPS charges go, the $H$ flux inside a 5-brane
is identified in M-theory with a membrane flux.
This means that (after compactification)
as a result of a scattering of a membrane on a 5-brane
an $H$-flux can be created and the membrane can be annihilated.
The identification of the $H$-flux with the membrane charge
is also what allows a membrane to end on a 5-brane \cite{openp}.
Consistency implies that a 5-brane with an $H$ flux should 
exert the same force on other objects as a 5-brane and a membrane.
This is indeed the case, as follows from the $C_3\wdg H$ interaction
on the 5-brane world-volume \cite{openp}.

The Lorentz force acting on a point like particle equals (in its rest
frame)
  \eqn{lorfor}{m\frac{d^2}{dt^2} x^i=e\cdot F^{0i}.}
As a generalization for a force acting on the fivebrane because of the
flux $H$ in the other 5-brane, we can replace $d^2/{dt^2}$ by
$\Box$ and write
  \eqn{lorfiv}{\Box \PH_A = F_{A\a\b\g}H^{\a\b\g}.}
But we must calculate the four-form supergravity field strength at the
given point. Only components with one Latin index and three Greek indices
are important. We note that the electric field strength in the real
physical 3+1-dimensional electrostatics is proportional to
  \eqn{elstatf}{F_{0A}\propto \frac{r_A}{r^3}\propto \frac 1{r^2}.}
The power 3 denotes 3 transverse directions, $F$ contains all the indices
in which the ``worldvolume'' of the particle is stretched. As an
analogue for fivebrane stretched exactly in $012345$ directions,
\eqn{statf}{*F_{012345A}\propto \frac{\PH^{(ij)}_A}{|\PH_{(ij)}|^5}
\propto \frac 1{|\PH_{(ij)}|^4}.}
We wrote star because we interpret the fivebrane as the ``magnetic''
source. $F$ in (\ref{lorfiv}) has one Latin index and three Greek 
indices, so its Hodge dual has four Latin indices and three Greek indices.
$*F$ in (\ref{statf}) contains only one Latin index but
when the 5-branes are tilted by infinitesimal angles
$\partial_\g\PH_C$ we get also a contribution
to the desired component of $F$:
  \eqn{desired}{*F_{\a\b\g ABCD}=
*F_{\a\b\g\d\s\t D}
\partial^\d \PH^{(ij)}_A
\partial^\s \PH^{(ij)}_B
\partial^\t \PH^{(ij)}_C.}
Now if we substitute (\ref{statf}) to (\ref{desired}) and the result
insert to (\ref{lorfiv}), we get the desired form of the $\Box \PH$
equations.


Similarly, there is an equation for $\TH$,
\bear
\dslash
\TH^i &\propto&
\sum_{j=1\dots N}^{(j\neq i)}
\frac{\epsilon^{ABCDE}}{\abs{\PH^{(ij)}}^5}
(\PH^{(ij})_E
\partial_\a(\PH^{(ij)})_A
\partial_\b(\PH^{(ij)})_B
\partial_{\g}(\PH^{(ij)})_C
\Ga^{\a\b\g}\Ga_D\TH^{(ij)}
\nn\\
\label{eqtheta}
\eear


Our goal in this paper is to deduce the corrections in the derivative
expansion in the low-energy of the (2,0) theory.
We cannot automatically deduce that (\ref{eqh}), (\ref{eqphi})
and (\ref{eqtheta})
can be extrapolated to the (2,0) theory because this description
is valid only in the opposite limit, when $|\PH|\ll M_p$, and supergravity
is not a good approximation.
However, the RHS of (\ref{eqh}) is a closed 4-form on
the moduli space $\Modsp = \MR{5}/\IZ_2 - \{0\}$
which is also half integral, i.e. in 
$H_4(\Modsp,{1\over 2}\IZ)$.
It must remain half-integral as we make $|\PH|$
smaller. Otherwise, Dirac quantization will be violated.
(Note that the wrapping number is always even.)
Eqn. (\ref{eqphi}) follows from the same term in the action
as (\ref{eqh}). As for other correction terms,
if we can show that they are implied by (\ref{eqh}) and supersymmetry,
then we can trust them as well.
This will be the subject of the next section.

We would like to point out that this reasoning is somewhat
similar to that of \cite{GG,GGV} who related the $R^4$ terms in 11D
M-theory to the $C\wdg X_8(R)$ term of \cite{VWSCT,DLM}.

\section{Compactification}

In this section we will study the reduction of the terms to 3+1D by
compactifying on $\MT{2}$.
Let $\Ar$ be the area of $\MT{2}$ and $\tau$ be its complex structure.
At low-energy in 3+1D we obtain a free vector multiplet of $N=4$ with
coupling constant $\tau$. We are interested in the subleading corrections
to the Wilsonian action. We will study these corrections as a function of
$\Ar$. Let us first note a few facts (see \cite{SeiSXN} for a detailed
discussion).

When one reduces classically a free tensor multiplet from 5+1D down
to 3+1D one obtains a free vector-multiplet with one photon and
6 scalars. Out of the 6 scalars one is compact. This is the
scalar that was obtained from $B_{45}$. We denote it by $\sigma$.
$$
\sigma = (\Imx\tau)^{-1/2}\Ar^{-1/2}\int_\MT{2} B_{45}.
$$
We have normalized its kinetic energy so as to have an $\Imx\tau$
 in front, like  3+1D SYM.
The radius of $\sigma$ is given by,
\begin{equation}
\label{sigper}
\sigma\sim \sigma + 2\pi (\Imx\tau)^{-1/2}\Ar^{-1/2}.
\end{equation}
In 5+1D there was a $Spin(5)_R$ global symmetry.
$N=4$ SYM has $Spin(6)_R$ global symmetry but the dimensional
reduction of the (2,0)-theory has only $Spin(5)_R$.
Let us also denote by $\VEV$ the square root of sum of squares
of the VEV of the 5 scalars other than $\sigma$.

Now let us discuss the interacting theory.
When $\VEV\Ar\ll 1$ we can approximate the 3+1D theory at
energy  scales $E\ll \Ar^{-1}$ by 3+1D SYM. In this case the
$Spin(5)_R$ is enhanced, at low-energy, to $Spin(6)_R$.
 For $\VEV\Ar\gg 1$ the ``dynamics'' of the theory occurs at length
scales well below the area of the $\MT{2}$ where the theory is 
effectively (5+1)-dimensional.
The 3+1D low-energy is therefore the classical dimensional
reduction of the 5+1D low-energy. Thus, from our 3+1D results
below we will be able to read off the 5+1D effective low-energy
in this regime.

\subsection{Dimensional reduction of the correction term}
Let us see what term we expect to see at low-energy in 3+1D.
We take the term,
$$
\px{[\a}H_{\b\g\d]} = 
{3\over {16\pi |\PH|^5}}\epd{ABCDE}\PH^E
\px{\a}\PH^A\px{\b}\PH^B\px{\g}\PH^C\px{\d}\PH^D,
$$
and substitute $0123$ for $\a\b\g\d$. 
The field $H_{\b\g\d}$ is,
$$
H_{\b\g\d} = -\ept{\b\g\d}{\a}H_{\a 45}
= -(\Imx\tau)^{1/2}\Ar^{-1/2}\epd{\b\g\d\a}\qx{\a}\sigma.
$$
The equation becomes
$$
\qx{\u}\px{\u}\sigma = -{1\over {32\pi}}
(\Imx\tau)^{-1/2}\Ar^{1/2}
{1\over {|\PH|^5}}\epd{ABCDE}\PH^E
\px{\a}\PH^A\px{\b}\PH^B\px{\g}\PH^C\px{\d}\PH^D
\epsilon^{\a\b\g\d}.
$$
Here $\PH^A\dots\PH^E$ are the six-dimensional fields.
The 4-dimensional fields are defined by,
\begin{equation}
\label{sixfour}
\PH^A = (\Imx\tau)^{1/2}\Ar^{-1/2}\vph^A.
\end{equation}
Thus, the action should contain a piece of the form,
\bear
\lefteqn{
{1\over {32\pi}}(\Imx\tau)\int d^4 x\,\,\px{\u}\sigma\qx{\u}\sigma
}\nn\\&&
-{1\over {32\pi}}(\Imx\tau)^{1/2}\Ar^{1/2}\epd{ABCDE}
\int d^4 x\,\, {{\sigma}\over {|\vph|^5}} \epu{\a\b\g\d}
\vph^E\px{\a}\vph^A\px{\b}\vph^B\px{\g}\vph^C\px{\d}\vph^D.
\label{sigfd}
\eear
Note that this is the behavior we expect when 
$\VEV\Ar\gg 1$. When $\VEV\Ar\sim 1$ the approximation
of reducing the 5+1D effective action is no longer valid as
explained above.

Let us first see how to write such a term in an $N=1$ superfield
notation.
Let us take three chiral superfields, $\CHP$ and $\CHP^I$
($I=1,2$). We assume that
$$
\CHP = \scVEV + \delta\vph + i\sigma.
$$
$\sigma$ is the imaginary part of $\CHP$ and
$\scVEV$ is the VEV of the real part.
Below, the index $I$ of $\CHP^I$ is lowered and raised with the
anti-symmetric $\epd{IJ}$.

\subsection{Interpolation between 3+1D and 5+1D}
In the previous section we assumed that we are in the region
$\VEV A \gg 1$. This was the region where classical
dimensional reduction from 5+1D to 3+1D is a good approximation.
However, the question that we are asking about the low-energy effective
action makes sense for any $A$. For $\VEV A \sim 1$ quantum effects
are strong.
Let us concentrate on another possible 
term which appears in the 5+1D effective
action and behaves like,
\begin{equation}
\label{vfrt}
\int d^6 x{{(\partial\PH)^4}\over {|\PH|^3}}.
\end{equation}
This term is of the same order of magnitude as (\ref{sigfd})
and its existence in the 5+1D effective action is suggested
by M(atrix) theory. It would give the correct $v^4/r^3$ behavior
for the potential between far away gravitons in M-theory
compactified on $\MT{4}$. We will also see below how terms
similar in structure to  (\ref{vfrt}) are related to (\ref{sigfd})
by supersymmetry.

After dimensional reduction to 3+1D we obtain a term which
behaves like
\begin{equation}
\label{vfrtdr}
(\Imx\tau)^{1/2}\Ar^{1/2}
\int d^4 x{{(\partial\vph)^4}\over {|\vph|^3}}.
\end{equation}
This is valid when $\VEV\Ar \gg 1$. On the other hand, when
$\VEV\Ar \ll 1$, $N=4$ SYM with a coupling constant
given by the combination $\tau$ is a good approximation, at low enough
energies (around the scale of $\VEV \Ar^{1/2}$).
In SYM, 1-loop effects can produce a term that behaves like (see \cite{DS}),
\begin{equation}
\label{vsymonlp}
\int d^4 x {{(\partial\vph)^4}\over {|\vph|^4}}.
\end{equation}
Note that this term contains no $\tau$, and no $\Ar$.

How can we interpolate between (\ref{sigfd}) and (\ref{vsymonlp})?

The answer lies in the periodicity of $\sigma$.
For any value of $\VEV\Ar$ the formula must be periodic
in the 6th scalar $\sigma$, according to (\ref{sigper}).
Thus, we propose to write
\begin{equation}
\int d^4 x{(\partial\vph)^4} 
\sum_{k\in\IZ} {1\over 
 {[\sum_{A=1}^5|\vph_A|^2 
   + (\sigma + 2 k \pi (\Imx\tau)^{-1/2}\Ar^{-1/2})^2]^2}}.
\end{equation}
For small $\Ar$ we can keep only the term with $k=0$ and recover
(\ref{vsymonlp}). For large $\Ar$ we have to approximate
the sum by an integral and we obtain
\begin{eqnarray}
\lefteqn{
\sum_{k\in\IZ} {1\over 
 {[\sum_{A=1}^5|\vph_A|^2 
    + (\sigma + 2 k \pi (\Imx\tau)^{-1/2}\Ar^{-1/2})^2]^2}}
} \nn\\
&\sim&
\int_{-\infty}^\infty {{dk}\over
 {[\sum_{A=1}^5|\vph_A|^2 
     + (\sigma + 2 k \pi (\Imx\tau)^{-1/2}\Ar^{-1/2})^2]^2}}
 = {1\over 4}(\Imx\tau)^{1/2}\Ar^{1/2}
{1\over {(\sum_{A=1}^5|\vph_A|^2)^{3/2}}}.
\nn
\end{eqnarray}
Thus we recover roughly (\ref{vfrtdr}).
One can make a similar conjecture for the generalization of 
(\ref{sigfd}) by changing the power of the denominator in the denominator
from $2$ to $5/2$ and modifying the numerator according to 
(\ref{sigfd}).
It is also easy to see, by Poisson resummation, that the corrections
to the integral fall off exponentially like (using (\ref{sixfour})),
$$
exp\left\{-(\Imx\tau)^{1/2}\Ar^{1/2}(\sum_{A=1}^5|\vph_A|^2)^{1/2}\} \right\}
 = 
e^{-\VEV \Ar},
$$
and so are related to instantons made by strings wrapping the $\MT{2}$.
There are no corrections which behave like Yang-Mills instantons,
i.e. $e^{2\pi i\tau}$. The reason for this was explained
in \cite{DS},  in the SYM limit.

\subsection{A derivation from 4+1D SYM}
When we compactify the $(2,0)_{(N=2)}$ theory
on $\MS{1}$ of radius $L$, we find a low-energy
description of $U(1)^2$ SYM. When $\VEV L^2 \ll 1$ and when the
energies are much smaller than $L^{-1}$, the effective 4+1D
SYM Lagrangian with $U(2)$ gauge group is a good approximation.

The moduli space is $\MR{5}/\IZ_2$ and the term (\ref{eqh})
implies that there is a term in the Lagrangian which is 
proportional to (we have switched to physical units),
$$
g\epd{ABCDE}
\int d^5 x\,\, {{1}\over {|\vph|^5}} \epu{\a\b\g\d\u} A_{\u}
\vph^E\px{\a}\vph^A\px{\b}\vph^B\px{\g}\vph^C\px{\d}\vph^D.
$$
This term can actually be seen as a 1-loop effect!
Let us consider a loop of a charged gluino with
4 external legs of scalars and 1 external leg of a photon.
Let the external momenta be 
$$
k_1,k_2,\cdots, k_5
$$
The loop behaves as,
\begin{equation}
\label{fivloop}
g^{5} \tr{t^1  t^2  t^3  t^4}
\int d^5 p\,\, \tr{\gamma_\u
{1\over {\pslash - m}}
\cdot
{1\over {\pslash + \kslash_1 -  m}}
\cdots
{1\over {\pslash + \kslash_1 +\cdots \kslash_4 -  m}}}.
\end{equation}
Here $m$ is the mass of the gluino and is proportional
to $g\vph_0$.
The coupling constant $g$ is proportional to $\sqrt{L}$ (see appendix).
The term with $\epu{\a\b\g\d\u}$ comes from expanding (\ref{fivloop})
in the $\kslash_i$.
We find
$$
g^{5}
\tr{t^1 t^2 t^3 t^4}
 \tr{\gamma_\u\kslash_1\kslash_2\kslash_3\kslash_4}
m\int {{d^5 p}\over {(p^2 + m^2)^5}}
\sim
g^5 m^{-4} 
\epu{ABCDE}
\epd{\a\b\g\d\u} k_1^\a k_2^\b k_3^\g k_4^\d.
$$
This is the behavior that we want.
It would be interesting to check
if a similar term appears in the low energy description of
the M-theory on $T^6$ as a matrix model
\cite{whycorrect}-\cite{HL}.
In a certain regime we can approximate
by 6+1D Yang-Mills. For the $SU(2)$ case the moduli space is
$\IR^3/ \IZ_2$. A similar effect could generate
term of the form below.
$$
\int A\wdg F\wdg F\wdg \epsilon_{ABC}
{{\phi^A d\phi^B\wdg d\phi^C}\over {|\phi|^3}}.
$$
After completion of this work, we have found out that
such terms were indeed calculated in \cite{thomp}.
We are grateful to G. Thompson for pointing this out to us.

\subsection{Component form}
Let us see how to write the term (\ref{sigfd})
in an $N=1$ superfield notation.
Let us take three chiral superfields, $\CHP$ and $\CHP^I$
($I=1,2$). We assume that
$$
\CHP = \scVEV + \delta\vph + i\sigma.
$$
$\sigma$ is the imaginary part of $\CHP$ and
$\scVEV$ is the VEV of the real part.
Below, the index $I$ of $\CHP^I$ is lowered and raised with the
anti-symmetric $\epd{IJ}$.

Let us examine the following term
\bear
I_1 &=& 
{1\over {32\pi}}(\Imx\tau)^{1/2}\Ar^{1/2}
\int d^4 x d^4\theta
  {1\over {(\CHP\bCHP + \CHP^I\bCHP_I)^{3/2}}}
  \bD^\da\bCHP^I D^\a\CHP_I
  \sig^{\u}_{\a\da}\bCHP^J\px{\u}\CHP_J + c.c.
\nn\\&&
\label{ione}
\eear
We can expand
\bear
\lefteqn{
\int d^4 x d^4\theta
  {1\over {(\CHP\bCHP + \CHP^I\bCHP_I)^{3/2}}}
  \bD^\da\bCHP^I D^\a\CHP_I
  \sig^{\u}_{\a\da}\bCHP^J\px{\u}\CHP_J 
} \nn\\
&=&
{1\over {\scVEV^3}}  \int d^4 x d^4\theta \bD^\da\bCHP^I D^\a\CHP_I
  \sig^{\u}_{\a\da}\bCHP^J\px{\u}\CHP_J 
\\&&
-
{3\over {2\scVEV^4}}  \int d^4 x d^4\theta 
  (\CHP + \bCHP - 2\scVEV)
  \bD^\da\bCHP^I D^\a\CHP_I
  \sig^{\u}_{\a\da}\bCHP^J\px{\u}\CHP_J 
+\Ox{{1\over {\scVEV^5}}}
\nn
\eear
Let us denote
\bear
I_2 &=& {i\over {8\scVEV^3}}  \int d^4 x d^4\theta \bD^\da\bCHP^I D^\a\CHP_I
  \sig^{\u}_{\a\da}\bCHP^J\px{\u}\CHP_J + c.c.,
\\
I_3 &=& {i\over {8\scVEV^4}}  \int d^4 x d^4\theta\,\,
  (\CHP + \bCHP - 2\scVEV)
  \bD^\da\bCHP^I D^\a\CHP_I
  \sig^{\u}_{\a\da}\bCHP^J\px{\u}\CHP_J + c.c.,
\nn
\eear
Let us check the bosonic part of $I_1$.
We use $\sc$ and $\sc^I$ for the scalar
components of $\CHP$ and $\CHP^I$.
We will expand in inverse powers of $\scVEV$ and
keep only leading terms.

It is easy to see that $I_3$ contains the term
\be
 {1\over {\scVEV^4}} \int d^4 x\,\,
  \sigma \epu{\a\b\g\d}\px{\a}\sc^I\px{\b}\sc_I\px{\g}\bsc^J\px{\d}\bsc_J
\ee
At the order of $1/{\scVEV^3}$ there are a few more terms that
do not include $\CHP$. They are listed below.
\bear
J_1 &=&  {1\over {\scVEV^3}} \int d^4\theta
    \px{\u}\CHP^I\qx{\u}\CHP^J\bCHP_I\bCHP_J
     +   {1\over {\scVEV^3}} \int d^4\theta
    \px{\u}\bCHP^I\qx{\u}\bCHP^J\CHP_I\CHP_J,
\nn\\
J_2 &=& {1\over {\scVEV^3}} \int d^4\theta
    \px{\u}\CHP^I\qx{\u}\bCHP_I\CHP^J\bCHP_J,
\nn\\
J_3 &=& {1\over {\scVEV^3}} \int d^4\theta
    \px{\u}\CHP^I\qx{\u}\bCHP^J\CHP_{I}\bCHP_{J},
\\
J_4 &=& {1\over {\scVEV^3}} \int d^4\theta
    \sig^{\u}_{\a\da} D^\a\CHP^I\bD^\da\bCHP_I \px{\u}\CHP^J\bCHP_J,
\nn
\eear
We now write down the bosonic terms of the above,
\bear
J_3 
&=& {1\over {\scVEV^3}}\int d^4x\,\{
 \sc_I\qx{\u}\sc^I\qx{\v}\bsc_J\px{\u}\px{\v}\bsc^J
-\sc_I\px{\u}\px{\v}\sc^I\bsc_J\qx{\u}\qx{\v}\bsc^J
\}\nn\\
J_2 
&=& {1\over {2\scVEV^3}}\int d^4x\,\{
    -4\sc_I\px{\u}\px{\v}\sc_J\bsc^I\qx{\u}\qx{\v}\bsc^J
    +2\sc_I\px{\u}\px{\v}\sc_J\bsc^J\qx{\u}\qx{\v}\bsc^I 
\\&&
    -2\sc_I\px{\u}\px{\v}\sc_J\qx{\v}\bsc^I\qx{\u}\bsc^J
    -3\sc_I\sc_J\px{\u}\px{\v}\bsc^I\qx{\u}\qx{\v}\bsc^J
\nn\\&&
     +\sc_I\px{\u}\sc^I\px{\v}\bsc_J\qx{\u}\qx{\v}\bsc^J
\}\nn\\
J_1
&=& {1\over {2\scVEV^3}}\int d^4x\,\{
    6\sc_I\px{\u}\px{\v}\sc_J\qx{\u}\bsc^I\qx{\v}\bsc^J
   +3\sc_I\px{\u}\px{\v}\sc_J\bsc^J\qx{\u}\qx{\v}\bsc^I
\nn\\&&\qquad
   +3\sc_I\px{\u}\px{\v}\sc_J\bsc^I\qx{\u}\qx{\v}\bsc^J
    +\sc_I\sc_J\px{\u}\px{\v}\bsc^I\qx{\u}\qx{\v}\bsc^J
\}\nn\\
J_4
&=& {8\over {\scVEV^3}}\int d^4x\,\{
\px{\u}\bsc_I\px{\v}\bsc^I\qx{\u}\sc_J\qx{\v}\sc^J
+\px{\v}\bsc_J\qx{\v}\bsc^I\qx{\u}\sc_I\px{\u}\sc^J
-2\bsc_J\qx{\u}\bsc^I\qx{\v}\sc_I\px{\u}\px{\v}\sc^J
\}\nn
\eear
We will now check which combination has 
the following symmetry which is part
of $SO(5)$ and doesn't involve $\sc$ and $\bsc$,
\be
\delta \sc^I = \bsc^I.
\ee
We find
\bear
\delta J_1
&=& {1\over {\scVEV^3}} \int d^4x\,\{
    4\sc_I\bsc_J\px{\u}\qx{\v}\bsc^I\px{\v}\qx{\u}\bsc^J
   -4\sc_I\px{\u}\sc^I\px{\v}\sc_J\qx{\u}\qx{\v}\bsc^J
\}\nn\\
\delta J_2
&=& {1\over {2\scVEV^3}}\int d^4x\,\{
     5\sc_I\px{\u}\bsc^I\px{\v}\bsc_J\qx{\u}\qx{\v}\bsc^J
    -3\sc_I\bsc_J\px{\u}\px{\v}\bsc^I\qx{\u}\qx{\v}\bsc^J
    +8\sc_I\px{\v}\sc^I\px{\u}\sc_J\qx{\u}\qx{\v}\bsc^J
\}\nn\\
\delta J_3
&=& {1\over {\scVEV^3}}\int d^4x\,\{
    2\sc_I\px{\u}\bsc^I\px{\v}\bsc_J\qx{\u}\qx{\v}\bsc^J
   +2\sc_I\px{\u}\sc^I\px{\v}\sc_J\qx{\u}\qx{\v}\bsc^J
\}\\
\delta J_4
&=& {8\over {\scVEV^3}}\int d^4x\,\{
2\sc_I\px{\v}\bsc_J\px{\u}\bsc^I\qx{\u}\qx{\v}\bsc^J
+2\sc_I\bsc_J\px{\u}\px{\v}\bsc^I\qx{\u}\qx{\v}\bsc^J
\}\nn
\eear
This puts some restrictions on the possible
${1 / {\VEV^3}}$ term. 
\bear\lefteqn{\scVEV^3 \delta (C_1 J_1 + C_2 J_2 + C_3 J_3 + C_4 J_4)
}\\&=&
    (4C_1 - \frac 32 C_2 + 16 C_4)
\sc_I\bsc_J\px{\u}\qx{\v}\bsc^I\px{\v}\qx{\u}\bsc^J
   +(\frac 52 C_2 + 2C_3 + 16 C_4)
\sc_I\px{\u}\bsc^I\px{\v}\bsc_J\qx{\u}\qx{\v}\bsc^J
\nn\\&&   +(-4C_1 + 4C_2 + 2C_3)
       \sc_I\px{\v}\sc^I\px{\u}\sc_J\qx{\u}\qx{\v}\bsc^J
\nn\eear
We see that
$$\frac 52 C_2 + 2C_3 + 16C_4 = 0,\qquad
4C_1 - \frac 32 C_2 + 16C_4 = 0.
$$
Thus, we need to take the following $SO(4)$ invariant combination
$$C(3J_1+8J_2-10J_3)+C'(4J_1+8J_3-J_4)
$$
where $C,C'$ are undetermined. We have not checked
if one can extend it to a
supersymmetric and $SO(5)$ invariant
combination by including interactions with
$\CHP$ \cite{toappear}.
We thank Savdeep Sethi for discussions on this point.

\section{Conserved quantities}

We can check that the overall ``center of mass'' decouples.
We can write it as a conservation equation for the 
total dissolved membrane charge ($j_Z$), total transverse momentum
($j_{\PH}$) and kinematical supersymmetry ($j_\TH$):
\eqn{kincur}{
j_Z^{\alpha,\beta\gamma}=\frac1{2\pi}
\sum_{i=1}^N H_i^{\alpha\beta\gamma},\quad
j^{\alpha,A}_{\PH}=\frac1{2\pi}
\sum_{i=1}^N\partial^\alpha\PH_i^A,\quad
j^\alpha_\TH=\frac1{2\pi}\sum_{i=1}^N \Ga^\alpha\TH_i.
}
They are conserved simply because $\partial_\alpha j^\alpha$ gives
the sum over $i,j$ of the right hand sides of
 (\ref{eqh}, \ref{eqphi}, \ref{eqtheta})
but the summand is $ij$ antisymmetric. The charges are defined as the
integrals of the $\alpha=0$ (lowered index) components
\eqn{charin}{
Z^{IJ}=\int \frac{d^5\s}{2\pi} \sum_{i=1}^N H_0^{i,IJ},\quad P^A=\int
\frac{d^5\s}{2\pi}\sum_{i=1}^N\partial_0\PH_i^A,\quad Q^{KIN}=\int
\frac{d^5\s}{2\pi}\sum_{i=1}^N\Ga_0\TH_i.
}
 We use the terms ``dissolved membranes'' and
``thin membranes'' for membranes of M-theory with 0 or 1 directions
transverse to the fivebranes, respectively. The thin membrane charge
appears as a central charge in the supersymmetry algebra \cite{SWSIXD}.
The 
reason is that $\{Q,\bar Q\}$ in M-theory contains momenta, twobrane and
fivebrane charges. But in (2,0)  theory, only the generators with $\GPR
Q=Q$ i.e. $\bar Q \GPR=-\bar Q$ survive. So we see that $\{Q,\bar Q\}$ is
a matrix anticommuting with $\GPR$ (i.e. containing an odd number of Greek
indices). For momenta it means that only momenta inside the fivebrane
worldvolume appear on RHS of supersymmetry algebra because $\Ga_\u$
anticommutes with $\GPR$ while the transverse $\Ga_A$ commutes with
$\GPR$. 

Only membrane charges contain $\Ga_\u\Ga_A$ which anticommutes
with $\GPR$ while $\Ga_{\u\v}$ and $\Ga_{AB}$ commute with $\GPR$. This is
an explanation why the thin membranes (looking like strings) with one
direction transverse to the fivebrane occur on the RHS of the
supersymmetry algebra.
There are also 3-form central charges which appear with
$\Ga_{\u\v\s}\Ga_A$ in the SUSY algebra. These correspond to 
tensor fluxes of the 3-form $H$ (analogous to electric and magnetic fluxes
in Yang-Mills theories).
But let us return to the thin membranes. We should be able to find the
corresponding current. The answer is (up to an overall normalization)
\eqn{thincur}{
M_{\a,\b}^A=-\frac 1{12\pi}
\epsilon_{\a\b\g\d\e\zeta}\sum_{i=1}^N
\partial^\g(\PH^A_i H_i^{\d\e\zeta})=
\frac{1}{2\pi}
\sum_{i=1}^N \partial^\g(H^i_{\a\b\g}\PH^A_i)}
The conservation law $\partial^\a M_{\a\b}^A$ is a simple consequence
of $\alpha\gamma$ anti-symmetry of $\epsilon_{\a\b\g\d\e\zeta}$. It is
also easy to see that for a configuration containing a membrane, the total
integral $\int d^5\s M_{0I}^A=W_I\cdot \Delta\PH^A$ measures the membrane
charge. Here $W_I$ is the winding vector of the induced string and $\Delta
\PH^A$ is the asymptotic separation of the two fivebranes.

There must be also a current corresponding to the $SO(5)$ R-symmetry.
It is given by
\eqn{Rcur}{
R_\a^{AB}= {1\over {2\pi}}
\sum_{i=1}^N
\left(
2\PH_i^{[A}\partial_\a\PH_i^{B]}
-\frac i2\bTH_i\Ga_\a\Ga^{AB}\TH_i
\right) + \mbox{corrections}.}

It is also quite remarkable that the corrected equations
conserve the stress energy tensor known from free theory.
For the initial considerations, let us restrict our attention to the
bosonic part of the stress tensor and choose the sign so that
$T_{00}>0$ i.e. $T^{0}_{\,\,\,0}<0$. 
Ignoring the requirement of the vanishing trace (i.e. without
the second derivatives that we discuss below),
the bosonic part 
of our stress tensor is given by
\eqn{strtensor}{T_{\alpha\beta}^{try}= {1\over {2\pi}}
\sum_{i=1}^N
\left(\frac 14 H^i_{\alpha\gamma\delta}H_\beta^{i,\gamma\delta}+
\partial_\alpha \PH_A^i\partial_\beta\PH_A^i-
\frac 12\eta_{\alpha\beta}\partial_\gamma\PH_A^i
\partial^\gamma\PH_A^i
\right).}
%
Note that the $\PH$ part has nonzero trace. The divergence
of this symmetric tensor can be written as
\eqn{diverg}{
\partial^\alpha T^{try}_{\alpha\delta} = {1\over {2\pi}}
\sum_{i=1}^N\left(
\frac 12 H^i_{\delta\gamma\delta'}
(\partial_\alpha H^{i,\alpha\gamma\delta'})
+(\Box \PH_D^i)\partial_\delta \PH_D^i
\right).
}
If we substitute $\Box\PH_D$ from (\ref{eqphi}) and 
$\px{\a}H^{i,\a\g\d'}$ from (\ref{eqh}) we obtain
$\partial^\alpha T^{try}_{\alpha\delta} = 0$.

We should note one thing that could be confusing. In the M-theory
containing $N$ fivebranes, the stress tensor is not equal to zero
but rather to\footnote{The minus sign in (\ref{cosmostr}) is
because our choice of the spacelike metric and $T_{00}>0$.}
  \eqn{cosmostr}{T^{M}_{\alpha\beta}=-N\tau^{(5)}\eta_{\alpha\beta}
+ T_{\alpha\beta}}
where our $T_{\alpha\beta}$ is just a small correction to the infinite
first term given by the tension of the fivebrane $\tau^{(5)}$. The
first term is in the limit of (2,0) theory infinite because
$\tau^{(5)}$ is of order $l_{Planck}^{-6}$ and $l_{Planck}$ is much
smaller than a typical distance inside fivebranes studied by (2,0) theory.
Nevertheless, gravity in this limit decouples and thus the
``cosmological'' term in (\ref{cosmostr}) plays no role.

\subsection{Traceless stress tensor and supercurrent}

In this subsection, we exhibit a traceless version of the stress tensor
and the supercurrent. We will use the adjective ``traceless'' both for the
supercurrent $J_\alpha$ and the stress tensor $T_{\alpha\beta}$
which means that
  \eqn{tlesuper}{\Ga^\alpha J_\alpha=0,\quad
T_\alpha^\alpha=0.}
The supercurrent has positive chirality $(\GPR-1) J_\alpha=0$ -- it
means
that the total supercharges have positive chirality as well.
We will also require continuity for stress tensor and supercurrent.
  \eqn{cont}{\partial^\alpha J_\alpha=0,\quad
\partial^\alpha T_{\alpha\beta}=0.}
Our definition of the stress tensor will be finally
\begin{eqnarray}
T_{\alpha\beta} &=& {1\over {2\pi}} \left\{
\sum_{i=1}^N
\frac 14
H^i_{\alpha\gamma\delta}H_\beta^{i,\gamma\delta}+\label{st}\right.\\
&+&\sum_{i=1}^N
\frac 3{5}\left(\partial_\alpha \PH_A^i\partial_\beta\PH_A^i-
\frac 16\eta_{\alpha\beta}\partial_\gamma\PH_A^i
\partial^\gamma\PH_A^i\right)-
\frac 2{5}
\PH_A^i\left(\partial_\alpha\partial_\beta-\frac 16
\eta_{\alpha\beta}\Box\right)\PH_A^i+\nonumber\\
&+& \left.\sum_{i=1}^N 
\frac{(-i)}2
\bar\TH^i\left(
\Ga_{(\alpha}\partial_{\beta)}-\frac 16\eta_{\alpha\beta}
\dslash
\right)\TH^i
\right\}
\label{stfermi}
\end{eqnarray}
We fixed a normalization for $H,\PH,\TH$ in this equation. The factors
$1/6$ inside the parentheses guarantee the tracelessness while the
relative factor $-3/2$ between the parentheses ensures vanishing of the
dangerous terms in $\partial^\alpha T_{\alpha\beta}$
which cannot be expressed from the equations of motion, namely
$\partial^{\alpha}\PH\partial_{\alpha\beta}\PH$. The $H^2$ part of the
stress tensor is traceless identically.
An explicit calculation shows that for the divergence of the stress
tensor we get
\begin{eqnarray}
\partial^\alpha T_{\alpha\beta} &=& {1\over {2\pi}}
\sum_{i=1}^N
\left[
\frac 12(\partial^\alpha  
H^i_{\alpha\gamma\delta})H_\beta^{i,\gamma\delta}
+\frac 23 (\partial_\beta\PH_A^i)(\Box\PH^A_i)-
\frac 13\PH^i_A(\partial_\beta\Box
\PH_i^A)+\right.\label{divst}
\\
&+&
\frac{7i}{12}(\partial_\beta\bar\TH^i)(
\dslash
\TH^i)-
\left.\frac i4\bar\TH^i\Ga_\beta\Box\TH^i-
\frac i6\bar\TH^i\partial_\beta
\dslash
\TH^i\right]\nonumber
  \end{eqnarray}

A similar approach can be used for the supercurrent as well. Here also
the $H\TH$ part is traceless identically 
while for the other parts it
is ensured by the $1/6$ factors. The relative factor $-3/2$ between the
parentheses is again chosen to cancel the dangerous $\partial^\alpha \PH
\partial_\alpha \TH$ terms in $\partial^\alpha J_\alpha$.
Note that the structure of $J_\alpha$ mimics the form of $T_{\a\b}$.
\begin{eqnarray}
J_\alpha &=&  {1\over {24\pi}}\sum_{i=1}^N
H_i^{\beta\gamma\delta}\Ga_{\beta\gamma\delta}\Ga_\alpha\TH^i
+\label{suscur}\\
&+& 
{1\over {\pi}}
\sum_{i=1}^N \left[
\frac 35\left(\partial_\alpha \PH_A^i
-\frac 16 \Ga_\alpha
\dslash
\PH_A^i\right)\Ga_A\TH^i
-\frac 25\PH_A^i\left(\partial_\alpha
-\frac 16 \Ga_\alpha
\dslash
\right)
\Ga^A\TH^i
\right]
\nonumber
\end{eqnarray}
We can compute also a similar continuity equation for the
supercurrent as we did for the stress tensor. The result is
  \bear  
\partial^\a J_\a &=&
{1\over {4\pi}}\partial_\a H^{\a\b\g}\Ga_{\b\g}\TH
+
{{1}\over {24\pi}} H^{\b\g\d}\Ga_{\b\g\d}\dslash\TH \\
&+&
{1\over {2\pi}}
\left[(\Box\PH_A)\Ga^A\TH-
\frac 13\dslash\PH_A\Ga^A(\dslash\TH)
-\frac 23\PH_A\Ga^A(\Box\TH)\right]\nn
 \eear
Using the equations of motion and the integration by parts, the
Hamiltonian and the total supercharge
defined as
\eqn{hasus}{
\ham=\int d^5\s T_{00}, \qquad
Q=\int d^5\s J_0
}
can be easily expressed as
  \eqn{hafr}{
\ham= {1\over {2\pi}}
\int d^5\sigma\left(
\frac 12 \left(\Pi^2+(\nabla \PH)^2\right)
+\frac 1{12} H_{KLM}H^{KLM}
+\frac{(-i)}2
\bar\TH \Ga^J\partial_J\TH \right).
}
(we use conventions with
$\com{\Pi(x)}{\PH(y)} = -2\pi i \delta^{(5)}(x-y)$\newline
and $\{ \TH^s(x),\bTH_{s'}(y)
\}=\pi((1+\GPR)\Ga_0)^{s}{}_{s'}\delta^{(5)}(x-y)$)
and,
\eqn{susr}{
Q={1\over {2\pi}} \int d^5\s
\left(
\frac{1}{6} H^{IJK}\Ga_{IJK}\Ga_0\TH
-\partial_\b\PH_A\Ga^\b\Ga^0\Ga^A\TH\right).
}
For convenience, we can also easily compute
\eqn{susrb}{
\bar Q={1\over {2\pi}} \int d^5\s
\left(
{1\over 6}
\bTH\Ga^0\Ga_{IJK}H^{IJK}+
\bTH \Ga^0\Ga^\b\Ga^A\partial_\b\PH_A\right)}
using a simple identity
  \eqn{barlo}{\overline{(\TH\Ga_{\u_1}\dots\Ga_{\u_N})}
=(-1)^N \bTH\Ga_{\u_N}\dots\Ga_{\u_1}.}
Now we can consider the supersymmetry transformation.
A variation of a field F will be written as
  \eqn{varf}{\d F = [\eb Q, F] = [\bar Q \e,F].}
Note that $\eb Q$ is an antihermitean operator because the components of
$Q$ or $\e$ are hermitean anticommuting operators or numbers,
respectively. Using the canonical commutation relations
we can easily compute the variations of the fields.
\eqn{vth}{
\d \TH=-[\TH,\bar Q\e]=
\left(
\frac{1}{6}\Ga_{IJK}H^{IJK} +
\Ga^\b\Ga^A\partial_\b\PH_A\right)\e}
This agrees with the transformations written before.
This, together with the normalization of $\acom{Q}{\bar{Q}}$ is
how we determined the relative coefficients.
Similarly,
\eqn{vph}{
\d\PH_A=[\eb Q,\PH_A]=\eb[Q,\PH_A]
=\eb[{1\over {2\pi}}
\smallint d^5\s 
\Pi_B\Ga^B\TH,\PH_A]= 
-i 
\eb\Ga_A\TH,
}
which agrees with previous definitions.
A similar but more tedious calculation gives us
\eqn{vha}{
\d H_{IJK}=\frac i2
\eb\cdot \epsilon_{0J'K'IJK}
\Ga^{I'J'K'}\Ga_0\partial_{I'}\TH,
}
which also agrees with the previous definition.

Let us summarize some formulas that are useful in understanding the
commutator of two supersymmetry transformations:
\bear
\d f=[\eb Q,f]=[\bar Q\e,f],& & 
\partial_\a f = -i[P_\a,f],\qquad P^0=\ham>0\\
 \{Q_s,\bar Q^{s'}\}&=&-2P^\u ((1+\GPR)/2\cdot\Ga_\u)
{}_s{}^{s'} + \mbox{thin}\\
\Rightarrow \delta Q&=&-[Q,\bar Q\e]=2P^\u\Ga_\u\e+ \mbox{thin}\\
\d J_\a &=&-2 T_{\a\b}\Ga^{\b}\e + \mbox{thin}\\
(\d_1\d_2-\d_2\d_1)f
&=&[\eb_1 Q,[\eb_2 Q,f]]-[\eb_2 Q,[\eb_1 Q,f]]=[[\eb_1 Q,\eb_2 Q],f]\\
&=&[\eb_1^s\{Q_s,\bar Q^{s'}\}\e_{2,s'},f]=-2[P^\u\eb_1\Ga_\u\e_2,f]
+ \mbox{thin}\\
&=&-2i(\eb_1\Ga^\u\e_2)\partial_\u f+ \mbox{thin}
\eear

\section{Speculations over a fundamental formulation}

In this section we would like to speculate on whether a fundamental
formulation of the (2,0) theory can be constructed from
the equations we discussed above.
We warn the reader in advance that this section could cause
some gritting of teeth!
Of course, the correction terms are not renormalizable if treated
as ``fundamental'' but let us go on, anyway. Perhaps some hidden
symmetry makes them renormalizable after all?

The model has the following virtues.
\begin{itemize}
\item there are absolutely no new fields. We use only $N$ copies of the
field strength $H_{MNP}$, five scalars $\PH_A$ and the 16 component
fermion $\TH$. Because of that, restriction to $N$ copies for distant
fivebranes is almost manifest.

\item the string current automatically satisfies the quantization
condition as a right winding number. This is related to the fact
that our current is automatically conserved (obeys the continuity
equation) which is necessary to allow us to insert it to equation
$dH=J$ -- and it has the correct dimension $mass^4$.

\item the total charge (sum over $1...N$) vanishes. The string (membrane
   connecting fivebranes) brings correctly minus source to one fivebrane
   and plus source to the other which agrees with the fact that
   the oriented membrane is outgoing from one fivebrane and incoming to
   another fivebrane -- and with $e_i-e_j$ roots of $U(N)$

\item the model is symmetric with respect to the correct 
Ho\hacek rava-Witten symmetry \cite{howitten} that accompanies the
reflection
$\PH_A^i\to-\PH_A^i$ by changing sign of $C_{MNP}$ (i.e. of $H_{MNP}$).

\item string states are given by strange configuration
   of fivebranes so that the vector of direction between two $\PH$'s
   draws whole $S^4$ (surface of ball in $\IR^5$) if one moves in
   the 4    transverse directions of the string.

\item $U(N)$ is not manifest, it arises due to the string states -- 
  perhaps in analogy with  the way enhanced symmetries appear in string
  theory because of D-brane    bound states.
\end{itemize}

What does a string look like? It is a solution constant in the time
and in one spatial direction, with a given asymptotical value of
$\Delta\PH=\abs{\PH^i-\PH^j}$ in infinity. We can show that such a
solution will
have typical size of order $\Delta\PH^{-1/2}$ in order to
minimize the tension (energy per unit of length of the string).

The value of $\partial\PH$ is of order $\Delta\PH/s$, integral of its
square over the volume $s^4$ is of order $(s\Delta\PH)^2$. On the
contrary, such a topological charge makes the field $H$ to behave
like $1/r^3$ where $r$ is the distance from the center of the solution.
Therefore $H$ inside the solution is of order $1/s^3$ which means that
the contribution of $H^2$ to the tension is of order $s^4/s^6=1/s^2$.
The total tension $(s\Delta\PH)^2+1/s^2$ is minimal for
$s=(\Delta\PH)^{-1/2}$ and the tension is therefore of order
$\Delta\PH$.
The field $\PH$ tries to shrink the solution while $H$ attempts to blow
it up.
In the next section we will describe the solution more concretely.


\section{String-like solution of (2,0) theory}
We will try to describe the string-like solution of the bosonic part
of the equations, considering only the topological term of $dH$ and the
corresponding term in $\Box\PH$ equation.
The following discussion
is somewhat reminiscent of a related discussion in \cite{CRRU}
for the effect of
higher order derivative terms on monopole solutions in $N=2$ Yang-Mills
but our setting is different.


\subsection{A rough picture}
Our solution will be constant in $\s^0,\s^5$ coordinates but it will
depend on the four coordinates $\s^1,\s^2,\s^3,\s^4$.
We are looking for a solution that minimizes the energy.
If the size of the solution in these
four directions is of order $s$, then the ``electric'' field, going like
$1/r^3$, is of order $1/s^3$ inside the solution and therefore the integral
$d^4\s (H^2)$, proportional to the tension, is of order $s^4/(s^3)^2$.

On the contrary, for the asymptotic separation $\Delta\PH$ quantities
$\partial \PH$ are of order $\Delta\PH/s$ inside the typical size of the
solution and therefore the contribution to the tension
$d^4\s (\partial\PH)^2$ is of order $s^4(\Delta\PH/s)^2$.

Minimizing the total tension $1/s^2+s^2 \Delta\PH^2$ we get the typical
size $s=(\Delta\PH)^{-1/2}$ and the tension of order $\Delta\PH$. In this
reasoning, we used the energy known from the free theory because the
bosonic part of the interacting stress energy tensor equals the free
stress energy tensor. The fact that the solution corresponds to the
interacting theory (and not to the free theory) is related to the
different constraint for $(dH)_{IJKL}$. 

\subsection{The Ansatz}

We will consider $N=2$ case of the (2,0) theory, describing two
fivebranes. Our solution will correspond to the membrane stretched between
these two fivebranes. Denoting by $(1)$ and $(2)$ the two fivebranes, we
will assume $\PH_{(1)}=-\PH_{(2)}$, $H_{(1)}=-H_{(2)}$ and we denote
$\PH_{(1)}$ and $H_{(1)}$ simply as $\PH$ and $H$.

Our solution will be 
invariant under $SO(4)_D$ rotating spacetime
and the transverse directions together.
The variable
$$r=\sqrt{\s_1^2+\s_2^2+\s_3^2+\s_4^2}$$
measures the distance from the center of the solution. 
We choose
the asymptotic
separation to be in the 10th direction and we denote it as
$$\PH^{10}(\infty)=\frac 12\Delta\PH.$$

Now there is an arbitrariness in the identification of the coordinates
$1,2,3,4$ and $6,7,8,9$. So there is in fact a moduli space of classical
solutions, corresponding to the chosen identification of these
coordinates. According to our Ansatz, the solution will be determined in
the terms of the three functions.
$$\PH^{I+5}=\s^If_1(r),\,\,I=1,2,3,4\quad
\PH^{10}=f_2(r),\quad
B_{05}=f_3(r).$$
We set the other components of $B_{\u\v}$ to zero and define $H$ as the
\asd part of $dB$,
$$H_{\a\b\g}=\frac 32 \partial_{[\a}B_{\b\g]} - \mbox{dual expression}.$$
It means that $H_{05I}=1/2\cdot\partial_I f_3$ and the selfduality
says
$$
H_{051}=-H_{234},\quad
H_{052}=H_{134},\quad
H_{053}=-H_{124},\quad
H_{054}=H_{123}.$$
Now we can go through the equations. $dH$ equations for $1,2,3,4$
determines $-4\partial_{[1}H_{234]}=\partial_IH_{05I}=\frac 12\Delta f_3$
where we used $\Delta=\Box$ because of the static character.
Therefore $dH$ equation says 
$$\Delta f_3=-8c_1\frac{f_1^3}{(f_2^2+r^2f_1^2)^{5/2}}
(-rf_1f'_2+f_1f_2+rf'_1f_2).$$
The three factors $f_1$ arose from $\partial_2\PH_7$,
$\partial_3\PH_8$,
$\partial_4\PH_9$,
we calculated everything at $\sigma^{1,2,3,4}=(r,0,0,0).$
At this point, only $EABCD=10,6789$ and $6,10,789$ from
$\epsilon$ symbol contributed.
Here $\Delta$ always denotes the spherically symmetric
part of the laplacian in 4 dimensions, i.e. 
$$\Delta=\frac{\partial^2}{\partial r^2}+\frac 3r\frac{\partial}{\partial
r}.$$
Similarly, we get hopefully two equations from $\Box\PH$. For $\PH^{10}$
(in the direction of asymptotic separation), we seem to get
$$\Delta f_2=6\frac{c_2}2\partial_1(-B_{05})
\frac{\e^{7,8,9,10,6}}{(f_2^2+r^2f_1^2)^{5/2}}rf_1^4.$$
Similarly, for the four other components we have
$$\Delta(rf_1)=-3c_2f'_3\frac{f_2 f_1^3}{(f_2^2+r^2f_1^2)^{5/2}}.$$

\subsection{Numerical solution, tension and speculations}

The functions $f_1,f_2,f_3$ are all even, therefore their derivatives
are equal to zero for $r=0$. The value of $f_2(0)$ finally determines
$f_2(\infty)$ which we interpret as $\Delta\PH /2$. The value of $f_1(0)$
must be fixed to achieve a good behavior at infinity and
$f_3(0)$ has no physical meaning, because only derivatives of $f_3=B_{05}$
enter the equations.

We can calculate the tension and we can compare the result
with the BPS formula. If we understand our equations just as some low
energy
approximation, there should be no reasons to expect that the calculated
tension will be precise, because
the approximation breaks down at the core. 

The tension expected from SYM theory is something like
$$M_W/L_5=\Delta\PH^{SYM}\cdot g/L_5=\sqrt{2\pi/L_5}\Delta\PH^{SYM}
=\sqrt{2\pi}\Delta\PH^{(2,0)}.$$
We just used simple formula for W boson masses, W bosons are string
wound around 5th direction and the $\PH$ fields of SYM and (2,0) are
related by $\sqrt{L_5}$ ratio as well.

The tension from our (2,0) theory is just twice (the same contribution
from two fivebranes) the integral
$$2\int d^4\s \frac 1{4\pi}\left(H_{05I}^2+(\partial_I\PH^A)^2\right)$$
Because of the spherical symmetry, we can replace $\int d^4\s$
by $\int_0^\infty dr\cdot 2\pi^2 r^3$. Work is in progress.

\section{Discussion}

Recently, a prescription for answering questions about
the large $N$ limit of the (2,0) theory has been proposed
\cite{Juan}.  In particular, the low-energy effective
description for a single 5-brane separated from $N$
5-branes has been deduced \cite{Juan}.
The topological term that we have discussed is, of course,
manifestly there. This is because a 5-brane probe in an
$AdS_7\times S^4$ feels the 4-form flux on $S^4$ and
and this will induce the anomalous $dH$ term.

What does M(atrix) theory have to say about non-linear
corrections to the low-energy of the (2,0) theory?
This is a two-sided question as the (2,0) theory
is a M(atrix) model for M-theory on $\MT{4}$ 
\cite{rozali,berozaliseiberg} and has a M(atrix) model
of its own \cite{prem,WitQHB}.

In order to be able to apply our discussion of the uncompactified
5+1D (2,0) theory to the M(atrix) model for M-theory
on $\MT{4}$ we need to be in a regime such that the VEV of
the tensor multiplet is much larger than the size of
$\MHT{5}$. This means that for a scattering process of
two gravitons in M-theory on $\MT{4}$ the distance between
the gravitons must remain much larger than the compactification
scale which we assume is of the order of the 11D Planck scale.
In this regime we expect the potential to behave
as $v^4/r^3$ (in analogy with $v^4/r^7$ in 11D).
Thus, things would work nicely if there were a term,
\begin{equation}
\label{phfour}
{{(\partial\PH)^4}\over {|\PH|^3}}
\end{equation}
in the effective low-energy description in 5+1D.
In the large $N$ limit, the existence of this term has been observed
in \cite{Juan}.
The term (\ref{phfour}) will also be the leading term
in the amplitude for a low-energy scattering of two
massless particles in the (2,0) theory.
It should thus be possible to calculate it from the M(atrix)
model of the (2,0) theory, with a VEV turned on.

It is also interesting to ask whether a term like (\ref{phfour})
is renormalized or not. An analysis which addresses such a question
in 0+1D will appear in \cite{PSS}. Perhaps a similar analysis
in 5+1D would settle this question.

\acknowledgments

We are grateful to
Micha Berkooz, Rami Entin, Savdeep Sethi, 
Zuzana Sakov\'a,
Herman Verlinde, Edward Witten,
and especially Tom Banks for valuable discussions.
The work of L. Motl was supported in part by the DOE under
grant number DE-FG02-96ER40559.
The work of O. Ganor was supported in part by the DOE
under the grant number DE-FG02-91ER40671.

\appendix
\section{Formulas for SUSY transformations}

In this text, we will use
the $SO(10,1)$ formalism for spinors, inherited from the M-theory
containing $N$ fivebranes, and
the space-like metric (in 5, 6 and 11 dimensions)
\bear
\eta_{\mu\nu} &=& \mbox{diag}(-++++++++++), \quad
\mu,\nu=0,1,\dots 10.\label{spacelike}\\
ds^2 &=& \eta_{\alpha\beta}dx^{\alpha}dx^{\beta} = -dx_0^2
+dx_1^2+dx_2^2+dx_3^2+dx_4^2+dx_5^2\nonumber
\eear

\subsection{SUSY transformation}
The SUSY transformations of the free tensor multiplet in 5+1D
is given by,
\begin{eqnarray}
\delta H_{\a\b\g} &=& -\frac{i}{2} \bar\epsilon
                       \Ga_\d\Ga_{\a\b\g}\partial^\d\TH
                   =
                    -3 i \eb \Ga_{\lbr\a\b}\partial_{\g\rbr}\TH
                    + \frac{i}{2} \eb\Ga_{\a\b\g}\Ga_{\d}\partial^\d\TH
\nn\\
\delta \PH_A      &=& -i \bar\epsilon\Ga_A\TH         
\nn\\
\delta \TH        &=& (\frac{1}{12} H_{\a\b\g}
                 \Ga^{\a\b\g}+ \Ga^\a\partial_\a
                 \PH_A\Ga^A)\epsilon                 
\nn
\end{eqnarray}
Since we are dealing with corrections to the low-energy
equations of motion, it is important to keep terms which vanish
by the equations of motion.
The SUSY commutators are thus given by,
\begin{eqnarray}
(\delta_1\delta_2 -\delta_2\delta_1) H_{\a\b\g} &=&
   - 2 i (\eb_1\Ga^{\u}\e_2) \partial_{\u} H_{\a\b\g}
\nn\\
&& - \frac{i}{2}
\ept{[\a\b}{\d\a'\b'\g'}\eb_1\Ga_{\g]}\e_2\partial_{[\d}H_{\a'\b'\g']}
+4 i (\eb_1\Ga^{\d}\e_2) \partial_{[\d}H_{\a\b\g]}
\nn\\
(\delta_1\delta_2-\delta_2\delta_1)\TH &=&
   - 2 i (\eb_1\Ga^\u\e_2)\partial_{\u}\TH
\nn\\
&& -\frac{i}{24} \left\{ 18(\eb_2\Ga_\u\e_1)\Ga^\u
  - 6(\eb_2\Ga_\u\Ga_{A}\e_1)\Ga^\u\Ga^{A}
\right\} \Ga_\b\partial^{\b}\TH
\nn\\
(\delta_1\delta_2 - \delta_2\delta_1) \PH_A &=&
   -2 i (\eb_1\Ga^{\u}\e_2)\partial_{\u}\PH_A,
\nn
\end{eqnarray}
The equations of motion transform according to,
\begin{eqnarray}
\delta\left(\Ga_{\d}\partial^\d\TH\right) &=&
     \frac{1}{6} \Ga^{\d\a\b\g}\e\partial_{[\d}H_{\a\b\g]}
     + \Ga^A\e\partial_{\a}\partial^{\a}\PH_A,
\nn\\
\delta\left(\partial_{[\u}H_{\a\b\g]}\right) &=&
     \frac{i}{2}
     \eb\Ga_{[\a\b\g}\partial_{\u]}\left(\Ga_\d\partial^\d\TH\right),
\nn\\
\delta\left(\partial_\u\partial^\u\PH_A\right) &=&
     -i \eb\Ga_A\left(\Ga_{\d'}\partial^{\d'}\right)
     \Ga_\d\partial^\d\TH=-i\eb\Ga_A\Box\TH.
\nn
\end{eqnarray}

\section{Quantization}

The quantization of the free tensor multiplet was discussed 
at length in \cite{WitFBE}.
There is no problem with the  fermions $\TH$ and bosons $\PH^A$,
but the tensor field is self-dual and thus has to be quantized
similarly to a chiral boson in 1+1D.
This means that we second-quantize a free tensor field
without any self-duality constraints and then set to zero
all the oscillators with self-dual polarizations.

The analogy with chiral bosons is made more explicit if
we compactify on $\MT{4}$ and take the low-energy limit
we we can neglect Kaluza-Klein states. We obtain a 1+1D
conformal theory. This theory is described by compact chiral
bosons on a $(3,3)$ lattice. This is the lattice of
fluxes on $\MT{4}$. For $\MT{4}$ which is a product of
four circles with radii $L_i$ ($i=1\dots 4$),  we get
3 non-chiral compact bosons with radii
$$
{{L_1 L_2}\over {L_3 L_4}},\,\,
{{L_1 L_3}\over {L_2 L_4}},\,\,
{{L_1 L_4}\over {L_2 L_3}},
$$
Of course, in 1+1D, T-duality can replace each radius $R$ with
$1/R$ and thus $SL(4,\IZ)$ invariance is preserved.

If we further compactify on $\MT{5}$ the zero modes
will be described by quantum mechanics on $\MT{10}$, where
$\MT{10}$ is the unit cell of the lattice of fluxes.

\subsection{Commutators}

Let us write down the commutation relations.

We want to reproduce the equations of motion by the Heisenberg equations
\eqn{heis}{\partial_0 (L) = i[\ham,L]
\mbox{\quad where\quad} \ham=\int d^5\sigma T_{00}.}
We should be allowed to substitute $H,\PH,\TH$ for the operator $L$.
In the following text we will use indices $I,J,K,\dots$ for the spatial
coordinates inside the
fivebrane. We will keep the spacelike metric and the convention
  \eqn{levicc}{\epsilon_{12345}=\epsilon^{12345}=1.}

We have the equations $H=-*H$ and $dH=0$. Among the fifteen equations 
for the vanishing four-form $dH=0$ we find ten equations with
index 0. These will be satisfied as the Heisenberg equations 
(\ref{heis}). Remaining five equations with space-like indices will only
play a role of some constraints that are necessary for consistent
quantization as we will see.
Let us take the example of equations of motion for $(dH)_{0345}$.
\eqn{dh}{
0=\partial_0H_{345}
-\partial_3 H_{450}+\partial_4 H_{503}-\partial_5H_{034}=
\partial_0 H_{345}+\partial_3H_{123}
+\partial_4H_{124}+\partial_5H_{125}.
}
It means that we should have the commutator
\eqn{ihcom}{
i[\ham,H_{345}(\sigma')]=
-\partial^{(\sigma')}_IH_{12I}(\sigma'),
}
where the important part of hamiltonian is
\eqn{hham}{
\ham_H= {1\over {8\pi}}\int d^5\sigma 
H_{0IJ}H_0^{\,\,IJ}=\int d^5\sigma
\frac {1}{24\pi} H_{KLM}H^{KLM}.
}
But it is straightforward to see that the relation (\ref{ihcom})
will be satisfied if the commutator of $H$'s will be
\eqn{hcom}{
[H_{IJK}(\sigma),H_{LMN}(\sigma')]=
-6\pi i
\partial_{[I}^{(\sigma)}\delta^{(5)}(\sigma-\sigma')
\epsilon_{JK]LMN}.
}

What does all this mean for the particles of the $H$ field?
Let us study Fourier modes of $H$'s with $\pm p_I$ where 
$p_I=(0,0,0,0,p)$.
Then we can see that $H_{125}(p)=H_{125}(-p)^\dagger$ is
a dual variable to $H_{345}(-p)=H_{345}(p)^\dagger$ and similarly
for two other pairs which we get using cyclic permutations
$12,34\to 23,14\to 31,24$. So totally we have three physical polarizations
of the tensor particle (which is of course the same number like
that of polarizations of photon in $4+1$ dimensional gauge theory).

We can also easily see from (\ref{hcom})
that the $p$-momentum modes of variables
that do not contain index ``5'', namely
$H_{123},H_{124},H_{134},H_{234}$ commute with everything.
They (more precisely their $\partial_5$ derivatives)
exactly correspond to the components of $dH$,
namely 
  \eqn{dhcom}{(dH)_{1235},(dH)_{1245},(dH)_{1345},(dH)_{2345}}
that we keep to vanish as the constraint part of $dH=0$.
Let us just note that $(*_5dH)_I=0$ contains {\it four}
conditions only because $d(dH)=0$ is satisfied identically.
Anyhow, there are no quantum mechanical variables
coming from the components of $(dH)_I$. The
variables $dH$ are the generators of the two-form gauge
invariance
  \eqn{twoforinv}{B_{IJ}\mapsto B_{IJ}+\partial_I\lambda_J
-\partial_J\lambda_I.}
Note that for $\lambda_I=\partial_I\phi$ we get a trivial transformation
of $B$'s which is the counterpart of the identity $d(dH)=0$.

But what about the zero modes, the integrals of $H_{IJK}$ over
the five-dimensional space? These are the ten fluxes that should
be quantized, i.e. they should belong to a lattice. In the 4+1 dimensional
SYM theory they appear as four electric and six magnetic fluxes.
In the matrix model
of M-theory on $T^4$ these ten variables are interpreted as four compact
momenta and six transverse membrane charges.

The fact that ``unpaired'' degrees of freedom 
are restricted to a lattice
is an old story. For
instance, in the bosonic formulation of the heterotic string in 1+1
dimensions we have 16 left-moving 
(hermitean)
bosons (``\asd field strengths'')
$\alpha^i$, $i=1,\dots,16$ with commutation relations 
  \eqn{hetcom}{[\alpha^i(\sigma),\alpha^j(\sigma')]=i\delta'(\sigma
-\sigma')\delta^{ij}.}
After combining them to Fourier modes
  \eqn{hetfour}{\alpha^i(\sigma)=\sqrt{\frac 2\pi}\sum_{n\in\IZ}
\alpha^i_n e^{-2i\sigma n}
\quad\Leftrightarrow\quad
\alpha^i_n=\frac 1{\sqrt{2\pi}}\int_0^\pi \alpha^i(\sigma)
e^{2i\sigma n}d\sigma}
we get relations
  \eqn{hettc}{[\alpha^i_m,\alpha^j_n]=m\delta_{m+n}\delta^{ij},\quad
(\alpha^i_m)^\dagger=\alpha^i_{-m}}
and we can interpret $\alpha^i_n$ and $\alpha^i_{-n}$ 
for $n>0$ as annihilation
and creation operators respectively. The modes $\alpha^i_0$ are then
restricted to belong to a selfdual lattice. Roughly speaking,
$\alpha^i_0$ equals the total momentum and it equals to
the total winding vector due to selfduality -- but these two must
belong to mutually dual lattices. The lattice must be even
in order for the operator
  \eqn{lhet}{L=:\frac 12\sum_{n\in\IZ}
\alpha^i_{-m}\alpha^i_m: = :\frac 14\int_0^\pi \alpha(\sigma)^2d\sigma:}
to have integer eigenvalues. We see that the 480 ground level states
${\ket 0}_{\alpha^i_0}$
with $(\alpha_0^i)^2=2$ give the same value $L=1$ as the sixteen 
lowest excited states $\alpha^i_{-1}{\ket 0}_{\alpha_0=0}$. These combine
to the perfect number 496 of the states.

\subsection{Correspondence with Super Yang Mills}

We will use the normalization of the gauge theory with Lagrangian and
covariant derivative as follows
\eqn{symconv}{
 \Lag=-\frac 1{4g^2}F^{\mu\nu}F_{\mu\nu},\qquad
 D_\alpha=\partial_\alpha+iA_\alpha.
}
The hamiltonian for the $U(1)$ theory then can be written as
($i,j=1,2,3,4$)
\eqn{symham}{
\ham_{SYM}=
  \frac 1{2g^2}\int d^4\sigma
  \left[\sum_i (E_i)^2 + \sum_{i<j} (F_{ij})^2\right].
}
Let us consider compactification on a rectangular $T^5$ 
(the generalization for other tori is straightforward)
of volume $V= L_1 L_2 L_3 L_4 L_5$.
We should get (\ref{symham}) from our hamiltonian. Let us 
write $d^5\sigma$ as $L_5 d^4\sigma$ (we suppose that the fields are
constant in the extra fifth direction).
\eqn{osymham}{
  \ham_{SYM}^{(2,0)}=
  \frac {L_5}{4\pi}\int d^4\sigma\sum_{I<J<K} (H_{IJK})^2.
}
So it is obvious that we must identify (up to signs) 
$F_{\alpha\beta}$ with $H_{\alpha\beta5}\cdot g\sqrt{L_5/(2\pi)}$
e.g.
\eqn{symid}{
  H_{234}=\frac{E_1}{g\sqrt{L_5/(2\pi)}},\quad
  H_{125}=\frac{F_{12}}{g\sqrt{L_5/(2\pi)}}.
}
To change $A_i$ of the SYM theory by a constant, we must take the
phase $\phi$ of the gauge transformation to be a linear function
of coordinates. But it should
change by a multiple of $2\pi$ after we go around a circle. Thus
\eqn{phisym}{
  \phi=\frac{2\pi n_i}{L_i}\sigma_i,\quad
  A_i\to A_i+\frac{2\pi n_i}{L_i}.
}
The dual variable to the average value of $A_i$ is the integral
of $E_i/g^2$. We just showed that the average
value of $A_i$ lives on a circle with radius
and therefore $L_1L_2L_3L_4\cdot E_i/g^2$ belongs to the lattice
with spacing $L_i$. Similarly, we can obtain a nonzero magnetic flux
from the configuration
($A_i$ can change only by a multiple of the quantum in (\ref{phisym}))
\eqn{magfl}{
  A_i=\frac{2\pi n_{ij}}{L_i}\cdot \frac{\sigma_j}{L_j},
}
which gives the magnetic field
\eqn{magfi}{
  F_{ij}=\frac{2\pi n_{ij}}{L_i L_j}.
}
Therefore for the spacings of the average values of $E_i,F_{ij}$ we have
\eqn{fspac}{
\Delta E_i=\frac{g^2L_i}{L_1L_2L_3L_4},\quad
\Delta F_{ij}=\frac{2\pi}{L_i L_j}.
}  
Looking at (\ref{symid}) we can write for the averages of $H$'s e.g.
\eqn{hspac}{
\Delta H_{234}=
\frac{gL_1}{L_1L_2L_3L_4\sqrt{L_5/(2\pi)}},\quad
\Delta H_{125}=\frac{2\pi}{L_1L_2g\sqrt{L_5/(2\pi)}}
}
which can be extended to a six-dimensionally covariant form only
using the following precise relation between the coupling
constant and the circumference $L_5$
\eqn{radcou}{
g = \sqrt{2\pi L_5},
}
giving us the final answer for the spacing
\eqn{skok}{
\Delta H_{IJK}=\frac{2\pi}{L_I L_J L_K}.
}
The formula (\ref{skok}) can be also written as
\eqn{skokk}{
\frac 16  \oint H_{IJK}dV^{IJK}\in 2\pi\cdot \mathbb{Z},
}
or (using antiselfduality) as
\eqn{skokl}{
 \Delta \int d^5\sigma H^{0IJ}= 2\pi L^I L^J,
}
in accord with the interpretation of $H$ as the current of dissolved
membranes (the integral in (\ref{skokl}) is the total membrane charge).

\subsection{Normalization of the current}

We can also work out the value of $c_1$ in (\ref{eqh}).
Let us write this equation for $\alpha\beta\gamma\delta=1234$.
\eqn{dhnum}{
  \partial_{[1}H_{234]}
  =\frac 14(\partial_1 H_{234}
  -\partial_2 H_{341}+\partial_3 H_{412}-\partial_4 H_{123})
  =\frac 14\partial_\alpha H^{05\alpha}
  =J_{1234}}
We see from (\ref{symid}) that 
  \eqn{jone}{J_{1234}=\frac 1{4g\sqrt{L_5/(2\pi)}}
\sum_{i=1}^4 \partial_i E_i.}
The integral of $\partial_i E_i$ should be an integer
multiple of $g^2$ (in these conventions) and 
because of (\ref{jone}), the integral of $J_{1234}$
should be an integer multiple of
$\pi/2$ which was the way we determined the coefficient
in (\ref{eqh}).

\section{Identities}

\subsection{Identities for gamma matrices}
\begin{eqnarray}
\Ga^\a\Ga^\b &=& \Ga^{\a\b}+\eta^{\a\b}\\
\Ga^{\a'}\Ga^{\b'\g'}
&=&\Ga^{\a'\b'\g'} + \Ga^{\g'}\eta^{\b'\a'}
   - \Ga^{\b'}\eta^{\g'\a'},
\\
\Ga^{\b'\g'}\Ga^{\a'}
&=&\Ga^{\a'\b'\g'} - \Ga^{\g'}\eta^{\b'\a'}
   + \Ga^{\b'}\eta^{\g'\a'},
\end{eqnarray}

\be
\Ga^{\a'}\Ga^{\b'\g'}
+\Ga^{\b'\g'}\Ga^{\a'} = 2\Ga^{\a'\b'\g'}
\ee
\be
\Ga^{\a'}\Ga_{\a\b} -\Ga_{\a\b}\Ga^{\a'}
    = 4\del_{\lbr\a}^{\a'}\Ga_{\b\rbr}
\ee
\be
\Ga^\d\Ga^{\a\b\g} = \Ga^{\d\a\b\g} + 3\eta^{\d[\a}\Ga^{\b\g]}
\ee
\be
\Ga^\d\Ga^{\a\b\g} + \Ga^{\a\b\g}\Ga^\d
   = 6 \eta^{\d[\a}\Ga^{\b\g]}
\ee
\be
\Ga^{\a'\b'\g'} \Ga_{\a\b} -
\Ga_{\a\b} \Ga^{\a'\b'\g'}
=
  12 \Ga^{\d'\lbr\b'\g'}
    \del_{\lbr\a}^{\a'\rbr}\eta_{\b\rbr\d'}
\ee
\be
\Ga^{\a'\b'\g'} \Ga_{\a\b} +
\Ga_{\a\b} \Ga^{\a'\b'\g'}
=
  -12 \del_{\lbr\a}^{\lbr\a'}\del_{\b\rbr}^{\b'}
\Ga^{\g'\rbr}+2\Ga^{\a'\b'\g'}{}_{\a\b}
\ee
\be
\GPR\Ga_{\u_1\u_2\cdots\u_k}
=(-1)^{k(k+1)/2}
{1\over {(6-k)!}} \ept{\u_1\u_2\cdots\u_k}{\nu_1\nu_2\cdots\nu_{6-k}}
  \Ga_{\nu_1\nu_2\cdots\nu_{6-k}}
\ee
\be
\ept{\u_1\u_2\cdots\u_k}{\nu_1\nu_2\cdots\nu_{6-k}}
= (-1)^{k}
  {\epsilon^{\nu_1\nu_2\cdots\nu_{6-k}}}_{\u_1\u_2\cdots\u_k}
\ee
\be
\ept{\u_1\cdots\u_{6-k}}{\nu_1\cdots\nu_{k}}
         \Ga^{\u_1\cdots\u_{6-k}}
= (-1)^{k(k-1)/2}(6-k)!\cdot\GPR\Ga^{\nu_1\cdots\nu_k}
\ee
\be
\GPR^2 = +1
\ee
\begin{eqnarray}
\Ga^{\a\b}\Ga_\b &=& 5\Ga^\a\\
\Ga^{\a}\Ga_{\a} &=& 6\Id,\\
\Ga^{\a}\Ga_{\u}\Ga_{\a} &=& -4\Ga_{\u},\\
\Ga^{\a}\Ga_{\u_1\u_2\cdots\u_k}\Ga_{\a} &=&
       (-1)^k (6-2k)\Ga_{\u_1\u_2\cdots\u_k},\\
\Ga^{\a_1\cdots\a_l}\Ga_{\a_1\cdots\a_l} &=&
       {{6!}\over {(6-l)!}} (-1)^{l(l-1)/2}\Id,\\
\Ga^{\a _1\cdots\a_l}\Ga_{\u}\Ga_{\a_1\cdots\a_l} &=&
(-1)^{l(l+1)/2}{{(6-2l)5!}\over {(6-l)!}}\Ga_\u.
\end{eqnarray}
\begin{equation}
\begin{array}{rclcrcl}
\Ga^{\a\b}\Ga_{\a\b} &=& -30\Id,
& &\Ga^{\a\b\g}\Ga_{\a\b\g} &=& -120\Id,\\
\Ga^{\a\b}\Ga_\u\Ga_{\a\b} &=& -10\Ga_\u,
& &\Ga^{\a\b\g}\Ga_{\u}\Ga_{\a\b\g} &=& 0,\\
\Ga^{\a\b}\Ga_{\u\nu}\Ga_{\a\b} &=& 2\Ga_{\u\nu},
& &\Ga^{\a\b\g}\Ga_{\u\nu}\Ga_{\a\b\g} &=& 24\Ga_{\mu\nu},\\
\Ga^{\a\b}\Ga_{\u\nu\s}\Ga_{\a\b} &=& 6\Ga_{\u\nu\s},
& &\Ga^{\a\b\g}\Ga_{\u\nu\s}\Ga_{\a\b\g} &=& 0.
\end{array}
\end{equation}
Derivation for last equations:
\begin{equation}\begin{array}{rcl}
\Ga^{\a\b\g}\Ga_{\u\nu(\s)}\Ga_{\a\b\g} &=&
(\Ga^{\b\g}\Ga^{\a}
+ \Ga^{\g}\eta^{\b\a} - \Ga^{\b}\eta^{\g\a})
\Ga_{\u\nu(\s)}\Ga_{\a\b\g} =
\Ga^{\b\g}\Ga^{\a}\Ga_{\u\nu(\s)}\Ga_{\a\b\g}
\\&=&
\Ga^{\b\g}\Ga^{\a}\Ga_{\u\nu(\s)}
(\Ga_{\a}\Ga_{\b\g}+\Ga_\b\eta_{\a\g}-\Ga_\g\eta_{\a\b})
\\ 
&=& \Ga^{\b\g}\Ga^{\a}\Ga_{\u\nu(\s)}\Ga_{\a}\Ga_{\b\g} +
2 \Ga^{\g\b}\Ga_\b\Ga_{\u\nu(\s)}\Ga_\g.
\end{array}\end{equation}

\begin{equation}\begin{array}{rcl}
\Ga^A\Ga_A &=& 5\Id,\\
\Ga^A\Ga_B\Ga_A &=& -3\Ga_B,\\
\Ga^A\Ga_{BC}\Ga_A &=& \Ga_{BC},\\
\Ga^A\Ga_{B_1 B_2 \cdots B_k}\Ga_A &=&
(-1)^k (5-2k)\Ga_{B_1 B_2 \cdots B_k}.
\end{array}
\end{equation}

\be
\tr{\Ga^{\u_1\u_2\cdots\u_k}\Ga_{\nu_1\nu_2\cdots\nu_k}}
 = 32 k! (-1)^{{{k(k-1)}\over {2}}}
\del_{\lbr\nu_1}^{\lbr\u_1}\del_{\nu_2}^{\u_2} \cdots
\del_{\nu_k\rbr}^{\u_k\rbr}.
\ee
\be
\tr{\Ga^{\u_1\cdots\u_k}\Ga^{A_1 \cdots A_l}
    \Ga_{\nu_1\cdots\nu_k}\Ga_{B_1\cdots B_l}} 
 = 32 k! l! (-1)^{{{(k+l)(k+l-1)}\over {2}}}
\del_{\lbr\nu_1}^{\lbr\u_1}\del_{\nu_2}^{\u_2} \cdots
\del_{\nu_k\rbr}^{\u_k\rbr}   
\del_{\lbr B_1}^{\lbr A_1}\del_{B_2}^{A_2} \cdots
\del_{B_k\rbr}^{A_k\rbr}.
\ee

\subsection{Fierz rearrangments} 
We need an identity of the form,
\be
M_{mn} \equiv (\e_1)_m (\eb_2)_n
 = (\sum_{k=0}^6 \sum_{l=0}^2 C_{\u_1\cdots\u_k A_1\cdots A_l}
         \Ga^{\u_1\cdots\u_k}\Ga^{A_1\cdots A_l})_{mn}.
\ee
($l\le 2$ because $\Ga^{01\dots 10} = 1$.)
We then get:
\be
C_{\u_1\cdots\u_k A_1\cdots A_l}
 = {{(-1)^{{{(k+l)(k+l-1)}\over{2}}}}\over {32 k! l!}}
   \tr{M \Ga_{\u_1\cdots \u_k}\Ga_{A_1\cdots A_l}}.
\ee
Now we take
\be
\e_2 = -\GPR\e_2,\qquad
\e_1 = -\GPR\e_1.
\ee
and rearrange $M = \e_1\eb_2$.
Now $\GPR M = -M = -M \GPR$ and we see that only terms with
odd $k$ survive.
\begin{eqnarray}
M &\equiv& \e_1\eb_2
\\
 &=& \left(
-\frac{(\eb_2\Ga_\u\e_1)}{32}\Ga^\u
      +\frac{(\eb_2\Ga_\u\Ga_{A}\e_1)}{32}\Ga^\u\Ga^{A}
      +\frac{(\eb_2\Ga_\u\Ga_{AB}\e_1)}{64}\Ga^\u\Ga^{AB}\right)
     (1+\GPR)
\nn\\&&
      +{{1}\over {192}}(\eb_2\Ga_{\u\nu\s}\e_1)\Ga^{\u\nu\s}
      -{{1}\over {192}}(\eb_2\Ga_{\u\nu\s}\Ga_{A}\e_1)\Ga^{\u\nu\s}\Ga^A
      -{{1}\over
{384}}(\eb_2\Ga_{\u\nu\s}\Ga_{AB}\e_1)\Ga^{\u\nu\s}\Ga^{AB}.
\nn\\
N &\equiv& \e_1\eb_2 - \e_2\eb_1
\nn\\
 &=& \left(-{{1}\over {16}}(\eb_2\Ga_\u\e_1)\Ga^\u
      +{{1}\over {16}}(\eb_2\Ga_\u\Ga_{A}\e_1)\Ga^\u\Ga^{A}\right)
     (1+\GPR)
\\&&
-{{1}\over{192}}(\eb_2\Ga_{\u\nu\s}\Ga_{AB}\e_1)\Ga^{\u\nu\s}\Ga^{AB}.
\nn\\
L &\equiv& \e_1\eb_2 + \e_2\eb_1
\nn\\
 &=&
     {{1}\over {32}}(\eb_2\Ga_\u\Ga_{AB}\e_1)\Ga^\u\Ga^{AB}
     (1+\GPR)
\\&&
      +{{1}\over {96}}(\eb_2\Ga_{\u\nu\s}\e_1)\Ga^{\u\nu\s}
      -{{1}\over {96}}(\eb_2\Ga_{\u\nu\s}\Ga_{A}\e_1)\Ga^{\u\nu\s}\Ga^A.
\nn
\end{eqnarray}  
where we have used, e.g.
\be
\eb_2\Ga_{\u\nu\s}\e_1 = \eb_1\Ga_{\u\nu\s}\e_2.
\ee
For opposite chirality spinors we have to replace $\GPR$ by $-\GPR$.

For, perhaps, future use, we will also calculate this for
$M = \psi_1\eb_2$ with
\be
\e_2 = -\GPR\e_2,\qquad
\psi_1 = \GPR\psi_1.
\ee
\begin{eqnarray}
M &\equiv& \psi_1\eb_2
\\
 &=& \left(-{{1}\over {32}}(\eb_2\psi_1)\Id
      -{{1}\over {32}}(\eb_2\Ga_{A}\psi_1)\Ga^{A}
      +{{1}\over {64}}(\eb_2\Ga_{AB}\psi_1)\Ga^{AB}\right.
\nn\\&&\left.
      +{{1}\over {64}}(\eb_2\Ga_{\u\nu}\psi_1)\Ga^{\u\nu}
      +{{1}\over {64}}(\eb_2\Ga_{\u\nu}\Ga_{A}\psi_1)\Ga^{\u\nu}\Ga^A
      -{{1}\over {128}}(\eb_2\Ga_{\u\nu}\Ga_{AB}\psi_1)\Ga^{\u\nu}\Ga^{AB}
\right)
      (1+\GPR).\nn
\end{eqnarray}
we also need,
\begin{eqnarray}
N &\equiv& \Ga_{\a\b}\psi_1\eb_2\Ga^{\a\b}
\\
 &=& \left({{15}\over {16}}(\eb_2\psi_1)\Id
      +{{15}\over {16}}(\eb_2\Ga_{A}\psi_1)\Ga^{A}
      -{{15}\over {32}}(\eb_2\Ga_{AB}\psi_1)\Ga^{AB}\right.
\nn\\&&\left.
      +{{1}\over {32}}(\eb_2\Ga_{\u\nu}\psi_1)\Ga^{\u\nu}
      +{{1}\over {32}}(\eb_2\Ga_{\u\nu}\Ga_{A}\psi_1)\Ga^{\u\nu}\Ga^A
      -{{1}\over
{64}}(\eb_2\Ga_{\u\nu}\Ga_{AB}\psi_1)\Ga^{\u\nu}\Ga^{AB}\right)
      (1+\GPR).
\nn
\end{eqnarray}
and,
\begin{eqnarray}
K &\equiv& \Ga_A\psi_1\eb_2\Ga^A
\\
 &=& \left(-{{5}\over {32}}(\eb_2\psi_1)\Id
      +{{3}\over {32}}(\eb_2\Ga_{A}\psi_1)\Ga^{A}
      +{{1}\over {64}}(\eb_2\Ga_{AB}\psi_1)\Ga^{AB}  \right.
\nn\\&& \left.
      +{{5}\over {64}}(\eb_2\Ga_{\u\nu}\psi_1)\Ga^{\u\nu}
      -{{3}\over {64}}(\eb_2\Ga_{\u\nu}\Ga_{A}\psi_1)\Ga^{\u\nu}\Ga^A
      -{{1}\over {128}}(\eb_2\Ga_{\u\nu}\Ga_{AB}\psi_1)\Ga^{\u\nu}\Ga^{AB}
\right)
      (1+\GPR).
\nn
\end{eqnarray}
 
\subsection{Few notes about $\mfrak{spin}(5,1)$}
      
We use the eleven-dimensional language for the spinors. But nevertheless
one could be confused by some elementary facts concerning
the reality condition
for the spinor $(4,4)$ of $\mfrak{spin}(5,1)\times \mfrak{spin}(5)$.
The spinor
representation  
$4$ of $\mfrak{spin}(5)$ is quaternionic (pseudoreal). Therefore
$(4,4)$ of
$\mfrak{spin}(5)\times \mfrak{spin}(5)$ is a real 16-dimensional
representation.
But one might think that spinor $4$ of $\mfrak{spin}(5,1)$ is {\it
complex}
so that we cannot impose a reality condition for the $(4,4)$
representation.

But of course, this is not the case. The spinor representation $4$ of
$\mfrak{spin}(5,1)$ is {\it quaternionic} as well since the algebra
$\mfrak{spin}(5,1)$
can be
understood also as $\mfrak{sl}(2,\mathbb H)$ of $2\times 2$
quaternionic matrices with
unit determinant of its $8\times 8$ real form. This has the right
dimension
\eqn{fifteen}{4\cdot 4 - 1 = 15 = \frac{6\cdot 5}{2\cdot 1}.}
In the language of complex matrices, there is a matrix $j_1$ so that
\eqn{struct}{(j_1)^2=-1,\qquad  j_1M_1=\bar M_1 j_1}
for all $4\times 4$ complex matrices $M_1$ of $\mfrak{spin}(5,1)$. Of
course,
for the $4\times 4$ matrices $M_2$ in $\mfrak{spin}(5)$ there is also
such
a matrix $j_2$ that
\eqn{structwo}{(j_2)^2=-1,\qquad  j_2M_2=\bar M_2 j_2.}
An explicit form for the equations (\ref{struct}--\ref{structwo}) is
built from $2\times 2$ blocks
\eqn{struexam}{j_1=\tb{{rr}\circ&1\\-1&\circ},\qquad
M_1=\tb{{rr}\alpha&\beta\\ -\bar\beta&\bar\alpha}.}
In the $(4,4)$ representation of $\mfrak{spin}(5,1)\times
\mfrak{spin}(5)$ the
matrices
are given by $M=M_1\otimes M_2$ and therefore we can define a matrix $j$
that shows that $M$ is equivalent to a real matrix.
\eqn{realcond}{j=j_1\otimes j_2,\quad
j^2=1,\quad
jM=j_1M_1\otimes j_2M_2=\bar M_1j_1\otimes \bar M_2j_2=\bar M j.}
The algebra $\mfrak{spin}(5,1)$ is quite exceptional between the other
forms
of $\mfrak{spin}(6)$. The algebra $\mfrak{so}(6)$ is isomorphic to
$\mfrak{su}(4)$,
algebra $\mfrak{so}(4,2)$ to $\mfrak{su}(2,2)$ and algebra  
$\mfrak{so}(3,3)$ to $\mfrak{su}(3,1)$.
The other form of $\mfrak{su}(4)$ isomorphic to $\mfrak{so}(5,1)$ is
sometimes
denoted
$\mfrak{su}^*(4)$ but now we can write it as $\mfrak{sl}(2,\mathbb
H)$ as well
(the generators are
$2\times 2$ quaternionic matrices with vanishing real part of the trace).
>From the notation $\mfrak{sl}(2,\mathbb H)$ it is also obvious that
$\mfrak u(2,\mathbb H)=\mfrak{usp}(4)$ forms
a subgroup (which is isomorphic to $\mfrak{so}(5)$).


\end{document}